\documentclass[reprint,superscriptaddress,amsmath,amssymb,aps,prx]{revtex4-2}
\usepackage{graphicx}
\usepackage{dcolumn}
\usepackage{bm}
\usepackage{hyperref}
\usepackage{xcolor}
\usepackage{physics}

\begin{document}

\preprint{APS/123-QED}

\title{Banded non-Hermitian random matrices, neural networks, and eigenvalue degeneracies}
\author{Richard Huang}
\author{David R. Nelson}
\affiliation{Department of Physics, Harvard University, Cambridge, Massachusetts 02138, USA}

\date{\today}

\begin{abstract}
We study two-banded, non-Hermitian random matrices inspired by sparse neural networks with a circular, 1d topology. We focus on two paradigmatic models, an SSH chain and a ladder model, which have both non-Hermitian directional bias and random sign disorder in the hoppings. The random sign disorder, which follows Dale's Law, leads to localization of the eigenstates, while the directional bias drives a delocalization transition in these states. The competition between disorder and directional bias results in rich eigenspectra with loops of extended states in the complex plane surrounded by regions of localized ones, and the eigenvalues are all confined to an annular region. Furthermore, the distinct band structures of the SSH chain and ladder model lead to different delocalization phenomena. Even in the absence of disorder, tuning the directional bias can lead to an eigenvalue degeneracy, which is an exceptional point for the SSH chain but a diabolic point for the ladder. In the presence of the disorder, these special eigenvalue degeneracies are preserved and also highlight key stages in the delocalization process. For both models, increasing the directional bias initially delocalizes states starting from within the bands. For the SSH chain, for large enough directional bias, the delocalized states open up a hole in the spectrum in the complex plane, similar to prior results for single band systems. But for the ladder model, as the directional bias is increased, the states delocalize in two stages, leading to two separate loops of extended states with localized states in between. The precise contours on which the extended states reside can be predicted from the Lyapunov exponents associated with products of random transfer matrices, in agreement with direct numerical diagonalization. Although we focus on periodic boundaries, results are discussed for open boundaries as well.
\end{abstract}

\maketitle

\section{Introduction}

In the spirit of statistical physics, insights into complex systems can arise by modeling some aspects as random. Random matrix theory emerged as a useful approach to study problems such as the distribution of energy levels in heavy nuclei (as proposed by Wigner \cite{wigner1958distribution}). These systems led to a focus on random matrix ensembles that were \textit{dense} and \textit{Hermitian}. On the other hand, in the context of 1d Anderson localization, \textit{sparse, Hermitian} random matrices, as embodied in tight binding models, could be used to understand electronic properties of disordered solids, which were dominated by localized states \cite{anderson1958absence}. Inspired by flux line depinning in type-II superconductors, Hatano and Nelson found that a \textit{non-Hermitian} analog could lead to eigenfunction delocalization, even in one and two dimensions \cite{hatano1996localization, hatano1997vortex}. Non-Hermitian couplings also arise naturally in biological systems, such as neural networks \cite{sommers1988spectrum,rajan2006eigenvalue,amir2016non,zhang2019eigenvalue,tanaka2019non}, island population dynamics \cite{nelson1998non, shnerb1998winding, dahmen2007population}, and ecological networks \cite{allesina2015stability}. 

In the context of neural networks, two features we would like to focus on are Dale's Law and sparse connectivity. Dale's Law constrains neurons to be purely excitatory or inhibitory \cite{eccles1976electrical,rajan2006eigenvalue}. Sparse network connectivity can be motivated by ring attractor networks, such as the head direction cells in \textit{Drosophila melanogaster} \cite{kim2017ring, tanaka2019non}. It can also be motivated by a coarse-grained picture of neural networks with connectivity that decays over a length scale $l$, in which case clusters of neurons over a scale less than $l$ can be coarse-grained as a single site, with sparse connections in between \cite{amir2016non}.

Non-Hermitian matrices are also a mathematically rich subject, as the relaxing of the constraint of Hermiticity can lead to an wider set of phenomena. One notable example is an \textit{exceptional point}, which refers to when both the eigenvalues and eigenvectors of a non-Hermitian matrix coalesce \cite{kato1966perturbation, heiss2004exceptional, berry2004physics}; this is in contrast to a \textit{diabolic point} \cite{yarkony1996diabolical, ashida2020non}, in which only the eigenvalues coalesce. Exceptional points mark qualitative changes in behavior (for example, the critically damped harmonic oscillator \cite{landau1976mechanics}, the fluttering of flags \cite{argentina2005fluid}, and delocalization transitions in Hatano-Nelson models \cite{shnerb1998winding,dahmen2007population}), and they appear in contexts including, but not limited to, optics and photonics \cite{miri2019exceptional}, open quantum systems \cite{heiss1998collectivity}, and nonlinear and stochastic dynamics \cite{weis2025generalized}. In this paper, inspired by neural network models, we will investigate the connections between localization, non-Hermitian eigenvalue degeneracies, and multi-banded random matrices. 

\subsection{Neural networks and random matrix theory}

We will begin by introducing a simple model of a two-layer neural network with feedforward and recurrent connectivity, see \cite{dayan2005theoretical}. The firing rates of the neurons in the input layer (encoded in $\mathbf{u}$) and the rates for neurons in the output layer (encoded in the vector $\mathbf{v}$) are related according to 
\begin{equation}
    \tau \frac{d\mathbf{v}}{dt}=-\mathbf{v}+\mathbf{F}(W\mathbf{u}+M\mathbf{v}),
\end{equation}
where $\mathbf{h}=W\mathbf{u}$ is the total feedforward input to each neuron in the network, with the matrix element $W_{ij}$ representing the strength of the synapse from a neuron $j$ in the input layer to a neuron $i$ in the output layer. The matrix $M$ captures the weights of the synaptic connections within the output layer, with $M_{ij}$ representing the strength of the synapse from neuron $j$ to neuron $i$. Furthermore, according to Dale's law, the neurons are purely excitatory or inhibitory, so for each $j$, the matrix elements $M_{ij}$ have the same sign for all $i$ (i.e., each column has entries of the same sign) \cite{eccles1976electrical}. The relaxation timescale for the firing rate is set by $\tau$. The activation function $F$, often with a sigmoidal form, bounds the firing rates from above. A linearized model is 
\begin{equation} \label{eq:firing_rates}
    \tau \frac{d\mathbf{v}}{dt}=-\mathbf{v}+\mathbf{h}+M\mathbf{v},
\end{equation}
where the first, self-inhibition couplings are assumed to be the same for all neurons. The connectivity matrix $M$ likely has both structured and random components. In \cite{amir2016non}, a connectivity matrix corresponding to a tight-binding model was studied, with each site modeled as a coarse-graining of a dense cluster of connected neurons. A more complete model might replace each site with a neural cluster with dense, random internal connections, and so we are motivated to consider networks with a more complex spatial substructure. 

\subsection{Spatial substructure and two paradigmatic models}

In this paper, we will focus on disordered, non-Hermitian variants of systems with two bands: the SSH (Su, Schrieffer, Heeger) chain \cite{su1980soliton} and a ladder model. We can think of each neural cluster as a unit cell and treat the two-banded models as a first step towards understanding how sublattice degrees of freedom affect properties of sparse, non-Hermitian random matrices. Sparse random matrices with a block structure motivated by internal degrees of freedom have also been previously used in the context of other disordered systems, such as amorphous solids \cite{cicuta2018unifying}.

\begin{figure}
    \centering
    \includegraphics[width=\linewidth]{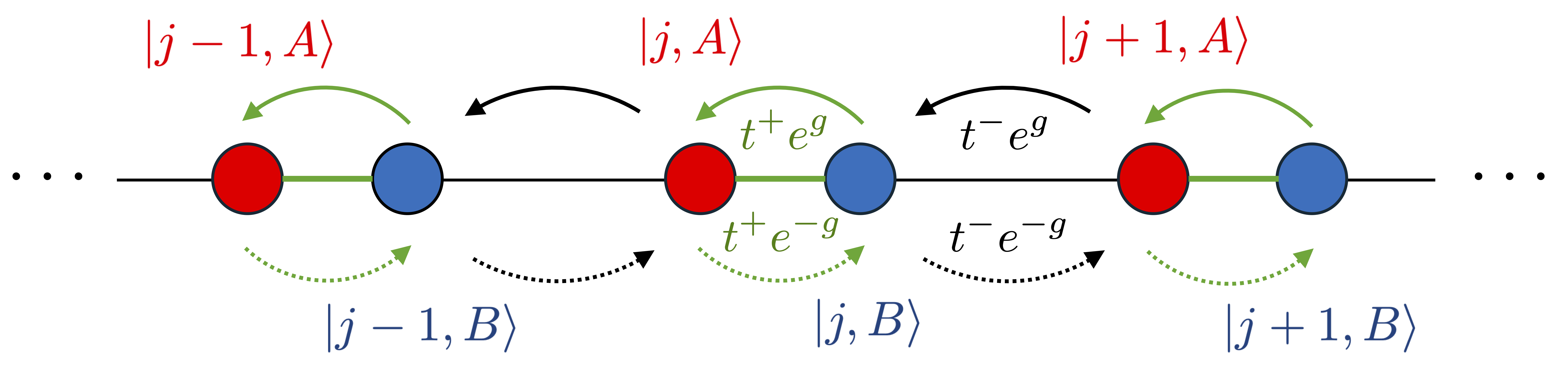}
    \caption{The non-Hermitian SSH chain. The unit cells are labeled $j=1,\cdots, N$, and each unit cell has an $A$ site (red) and $B$ site (blue). The intracell and intercell hopping parameters are $t^+=t+\delta t$ and $t^-=t-\delta t$, respectively. Furthermore, a parameter $g$ biases the hopping. Finally, we independently assign each site to be excitatory or inhibitory using a parameter $\sigma\in\{\pm1\}$, chosen to be $\pm 1$ with equal probability.}
    \label{fig:ssh_chain_setup}
\end{figure}

The first system we will consider is the non-Hermitian SSH chain, depicted in Figure \ref{fig:ssh_chain_setup}. 
The SSH chain comprises of $j=1,\cdots,N$ unit cells, each with two sites labeled by $A$ and $B$. There is hopping between adjacent sites, with $t^+=t+\delta t$ being the hopping strength within a cell and $t^-=t-\delta t$ being the hopping strength between adjacent cells. In addition, the hopping is not symmetric, with a parameter $g$ that biases the hopping. Finally, the hopping has random-sign disorder in accordance with Dale's law, which assumes that each neuron in a network has exclusively excitatory or inhibitory connections with its neighbors \cite{eccles1976electrical,dayan2005theoretical}: each site $(j,A/B)$ is independently chosen to be either excitatory ($\sigma_{j,A/B}=+1$) or inhibitory ($\sigma_{j,A/B}=-1$) with equal probability. The resulting connectivity matrix $M_S$ for this model is given by
\begin{equation} \label{eq:rs_ssh}
\begin{split}
    M_S &= t^+ \sum_{j=1}^N \Big[e^{g}\sigma_{j,B}\ket{j,A}\bra{j,B}+
    e^{-g}\sigma_{j,A}\ket{j,B}\bra{j,A}\Big] \\
    & + t^-\sum_{j=1}^N \Big[e^g\sigma_{j+1,A}\ket{j,B}\bra{j+1,A} \\ 
    & \hspace{3cm}+e^{-g}\sigma_{j,B}\ket{j+1,A}\bra{j,B})\Big],
\end{split}
\end{equation}
where for most of this paper we impose periodic boundary conditions: $\ket{j+N,A}=\ket{j,A}$ and $\ket{j+N,B}=\ket{j,B}$. In a basis of $\ket{1,A}, \ket{1,B},\ket{2,A},\cdots, \ket{N,B}$, we can represent the connectivity with the following sparse, banded $2N\times 2N$ matrix:
\begin{equation} \label{eq:rs_ssh_matrix}
    M_S= \begin{pmatrix}
        & p_2 & & & \ell_{2N}\\
        \ell_1 & & p_3 \\
        & \ell_2 & & \ddots \\
        & & \ddots & & p_{2N} \\
        p_1 & & & \ell_{2N-1}
    \end{pmatrix},
\end{equation}
where the non-zero entries are
\begin{equation}
    \begin{split}
        \ell_i &= \begin{cases}
        t^+e^{-g} \sigma_{\frac{i+1}{2}, A} \quad &(\text{for $i$ odd})\\
        t^-e^{-g} \sigma_{\frac{i}{2}, B} \quad &(\text{for $i$ even}),
    \end{cases} \\
    p_i &= \begin{cases}
        t^-e^g \sigma_{\frac{i+1}{2}, A} \quad &(\text{for $i$ odd})\\
        t^+e^g \sigma_{\frac{i}{2}, B} \quad &(\text{for $i$ even}).
    \end{cases}
    \end{split}
\end{equation}
Here, we have shifted the firing rates in Eq. (\ref{eq:firing_rates}) so that all diagonal elements of Eq. (\ref{eq:rs_ssh_matrix}) are zero. All other elements of $M_S$, except for those on the sub- and super-diagonal and also indicated at the corners, are zero as well.

\begin{figure}
    \centering
    \includegraphics[width=\linewidth]{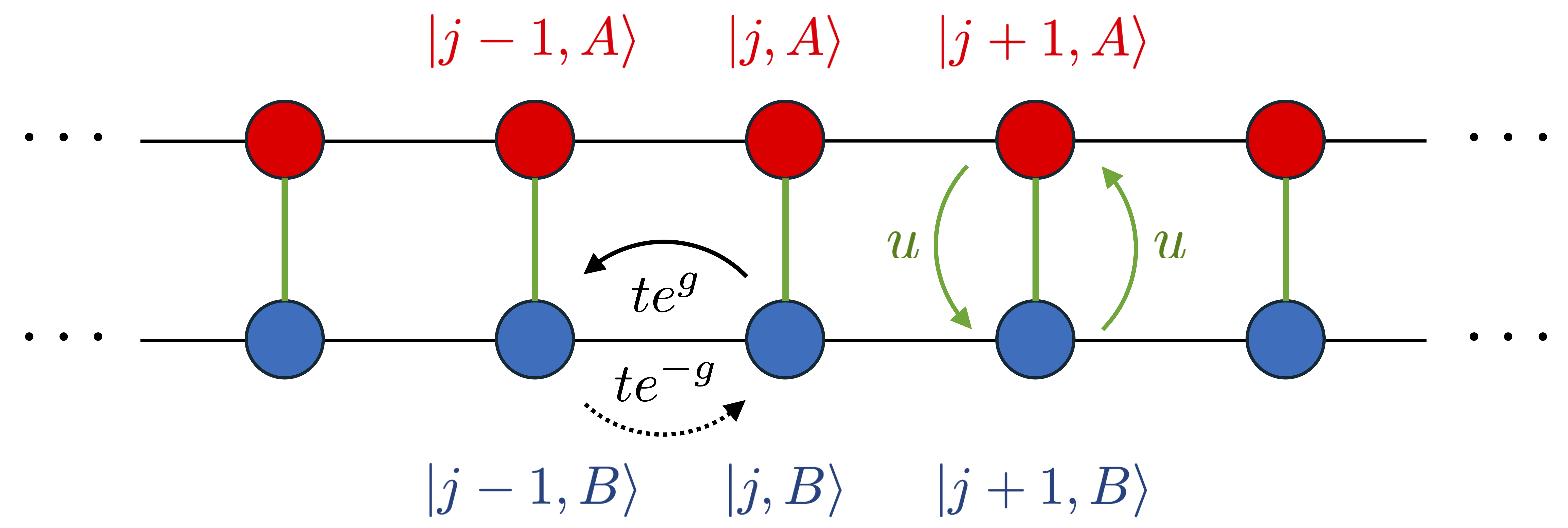}
    \caption{The non-Hermitian ladder. The unit cells are labeled $j=1,\cdots, N$, and each unit cell has an $A$ site (red) and $B$ site (blue). The intracell and intercell hopping parameters are $u$ and $t$, respectively. Furthermore, a parameter $g$ biases the hopping, but only along the upper and lower sublattices. Finally, we independently assign each site to be excitatory or inhibitory using a parameter $\sigma\in\{\pm1\}$, chosen to be $\pm 1$ with equal probability.}
    \label{fig:ladder_setup}
\end{figure}

To contrast with the non-Hermitian SSH chain, the second system we will consider is a non-Hermitian ladder model, depicted in Figure \ref{fig:ladder_setup}. 
The ladder comprises of $j=1,\cdots,N$ unit cells, each representing a ``rung" of the ladder with two sites $A$ and $B$. There is hopping $t$ between adjacent sites in the same sublattice, as well as hopping $u$ within a rung. The hopping within each sublattice is not symmetric, due to a biasing parameter $g$ along the ladder. However, the hopping within a rung remains symmetric. In other words, this ladder is comprised of two non-Hermitian Hatano-Nelson models with Hermitian couplings. Finally, we introduce random-sign disorder as before, with each site independently chosen to be excitatory or inhibitory with equal probability. The resulting connectivity matrix $M_L$ for this model is given by
\begin{equation} \label{eq:rs_ladder}
    \begin{split}
        M_L = t\sum_{j=1}^N \Big[e^g\sigma_{j+1, A}&\ket{j, A}\bra{j+1, A} \\
        &+ e^{-g}\sigma_{j, A}\ket{j+1, A}\bra{j, A}\Big]\\
        + t\sum_{j=1}^N \Big[e^g\sigma_{j+1, B}&\ket{j, B}\bra{j+1, B} \\
        &+ e^{-g}\sigma_{j, B}\ket{j+1, B}\bra{j, B}\Big]\\
        +u\sum_{j=1}^N\Big[\sigma_{j,B}\ket{j,A}&\bra{j,B}+\sigma_{j,A}\ket{j,B}\bra{j,A}\Big],
    \end{split}
\end{equation}
where for most of this paper we impose periodic boundary conditions: $\ket{j+N,A}=\ket{j,A}$ and $\ket{j+N,B}=\ket{j,B}$. In a basis of $\ket{1,A}, \ket{1,B},\ket{2,A},\cdots, \ket{N,B}$, we can represent the connectivity with the following sparse $2N\times 2N$ matrix:
\begin{equation} \label{eq:rs_ladder_matrix}
    M_L = \begin{pmatrix}
        R_1 & L_2^+ & & & L_N^-\\
        L_1^- & R_2 & L_3^+ \\
        & L_2^- & \ddots & \ddots \\
        & & \ddots & \ddots & L_N^+ \\
        L_1^+ & & & L_{N-1}^- & R_N
    \end{pmatrix},
\end{equation}
where $R_j$ and $L_j^\pm$ are the following $2\times 2$ matrix blocks:
\begin{equation} \label{eq:rs_ladder_blocks}
    R_j = u \begin{pmatrix}
        0 & \sigma_{j,B} \\
        \sigma_{j,A} & 0
    \end{pmatrix}, \quad
    L_j^\pm = te^{\pm g} \begin{pmatrix}
        \sigma_{j,A} & 0 \\
        0 & \sigma_{j,B}
    \end{pmatrix},
\end{equation}
and blocks not indicated explicitly in Eq. (\ref{eq:rs_ladder_matrix}) are zero.

\subsection{Outline}

The rest of the paper is organized as follows. In Section \ref{sec:nhssh}, we discuss the SSH chain. Even in the absence of random sign disorder in the hoppings, the non-Hermitian SSH chain admits an exceptional point. We will see how this exceptional point can remain even in the presence of random sign disorder, and investigate the gaps and symmetries in the random spectrum of eigenvalues in the complex plane. Increasing the directional bias eventually delocalizes the eigenstates, and we will show how the contours in the complex plane on which these extended states reside can be predicted within a transfer matrix formalism. In Section \ref{sec:nhladder}, we discuss analogous results for the ladder model. In particular, in the absence of disorder, the ladder model does not admit an exceptional point, but rather a diabolic point. In addition, the delocalization of the eigenstates with increasing bias parameter $g$ proceeds in a manner that is significantly different compared to that of the SSH chain. We will see how it occurs in two stages, leading to two distinct contours of extended states. In Section \ref{sec:obc}, we analyze the two systems for open boundary conditions. Finally, in Section \ref{sec:discussion}, we conclude with additional discussion and future directions.

\section{The non-Hermitian SSH chain} \label{sec:nhssh}

Before we consider the full model in Eq. (\ref{eq:rs_ssh}), let us remove the random sign disorder in the hopping parameters and briefly point out some of the interesting features of the resulting pure non-Hermitian SSH chain $H_S$:
\begin{equation} \label{eq:pure_ssh}
    \begin{split}
    H_S &= t^+ \sum_{j=1}^N (e^{g}\ket{j,A}\bra{j,B}+
    e^{-g}\ket{j,B}\bra{j,A}) \\
    &+ t^-\sum_{j=1}^N (e^g\ket{j,B}\bra{j+1,A}+e^{-g}\ket{j+1,A}\bra{j,B}).
\end{split}
\end{equation}
If we assume periodic boundary conditions $\ket{j+N,A}=\ket{j,A}$ and $\ket{j+N,B}=\ket{j,B}$, the system is translationally invariant and can be partially diagonalized using Bloch eigenfunctions $\ket{k,A}=\frac{1}{\sqrt{N}}\sum_j e^{ikj}\ket{j,A}$ and $\ket{k,B}=\frac{1}{\sqrt{N}}\sum_j e^{ikj}\ket{j,B}$ of the translation operator. For each momentum $k$ in the first Brillouin zone, we have the two level system
\begin{equation}
    H_S(k)=\begin{pmatrix}
        0 & t^+e^g+t^-e^{-(g+ik)} \\
        t^+e^{-g}+t^-e^{g+ik} & 0
    \end{pmatrix}.
\end{equation}
It is useful to decompose $H_S(k)$ in terms of the Pauli matrices $\bm{\sigma}=(\sigma_x,\sigma_y,\sigma_z)$: $H_S(k)=\mathbf{d}(k)\cdot\bm{\sigma}$. Since $H_S(k)$ is non-Hermitian, the vector $\mathbf{d}(k)$ will be complex, and it can be decomposed into its real and imaginary parts: $\mathbf{d}=\mathbf{d}_R+i\mathbf{d}_I$, where
\begin{equation}
    \begin{split}
        \mathbf{d}_R(k) &= \begin{pmatrix}
        t^++t^-\cos k \\ t^-\sin k \\ 0
    \end{pmatrix}\cosh g, \\
    \mathbf{d}_I(k) &= \begin{pmatrix}
        t^-\sin k \\ t^+-t^-\cos k \\ 0
    \end{pmatrix}\sinh g.
    \end{split}
\end{equation}
The eigenvalues of $H_S(k)$ can be expressed in terms of $\mathbf{d}_R$ and $\mathbf{d}_I$ \cite{bergholtz2021exceptional}:
\begin{equation} \label{eq:ssh_bands}
    E_\pm(k)=\pm\sqrt{d_R^2(k)-d_I^2(k)+2i\mathbf{d}_R(k)\cdot\mathbf{d}_I(k)}.
\end{equation}
If we consider the special case where $H_S$ is Hermitian ($g=0$), we get $\mathbf{d}_I=0$ so $E_\pm(k)=\pm|\mathbf{d}_R(k)|$. This recovers the standard SSH band structure
    \begin{equation}
        E_\pm(k)=\pm\sqrt{(t^+)^2+(t^-)^2+2t^+t^-\cos k},
    \end{equation}
and if $\delta t\neq 0$, there is always a band gap since $t^+\neq t^-$.

\begin{figure}
    \centering
    \includegraphics[width=\linewidth]{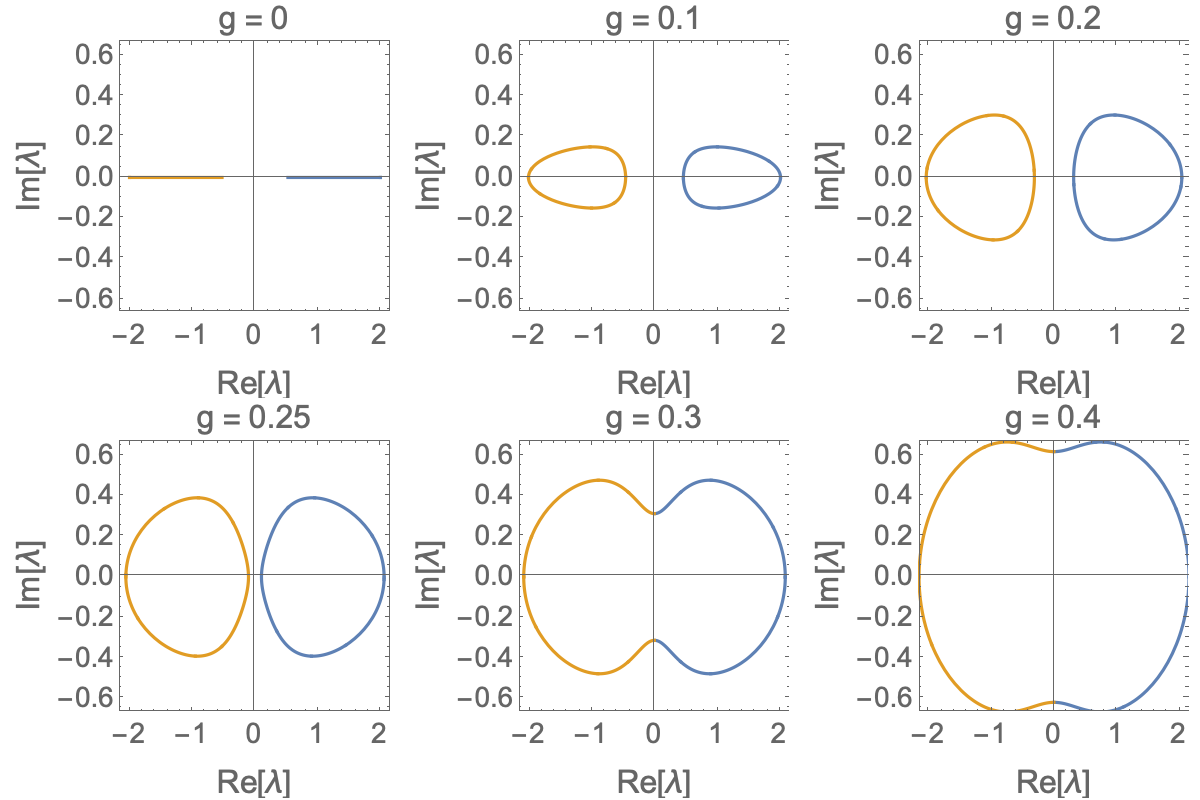}
    \caption{Eigenspectrum for the non-Hermitian SSH chain for $t=1$ and $\delta t=0.25$. As the non-Hermitian parameter $g$ is increased, eigenvalues wander off the real axis. The band gap closes at $g_c=\tanh^{-1}(\delta t/t)\approx 0.255$.}
    \label{fig:ssh_bands}
\end{figure}

However, the non-Hermitian case is different. Without loss of generality, we will focus on $g>0$. For $0<\delta t<t$, it is always possible to tune $g$ such that the point gap at $k=\pi$ closes. This is illustrated in Figure \ref{fig:ssh_bands}, where the SSH band structure from Eq. (\ref{eq:ssh_bands}) has been plotted for $t=1, \delta t=0.25$, and various values of $g$. 

The critical value of $g$ is given by
\begin{equation} \label{eq:ssh_gc}
    g_c=\tanh^{-1}\left(\frac{\delta t}{t}\right)=\frac{1}{2}\ln\left(\frac{t^+}{t^-}\right),
\end{equation}
for which $H_S(k=\pi)$ takes the form of a simple Jordan block:
\begin{equation}
    H_S(k=\pi; g=g_c)=\frac{(t^+)^2-(t^-)^2}{\sqrt{t^+t^-}}\begin{pmatrix}
        0 & 1 \\
        0 & 0
    \end{pmatrix}.
\end{equation}
This corresponds to an \textit{exceptional point}: there is a degenerate eigenvalue, but in contrast to Hermitian systems, there is an incomplete set of eigenvectors. In this case, the two-fold degenerate zero eigenvalue has only one right eigenvector: $\ket{\pi, A}$. In Figure \ref{fig:ssh_bands}, the exceptional point occurs for $g_c\approx 0.255$. In the following section, we will see how the exceptional point at $g=g_c$ persists even in the presence of disorder.

\subsection{SSH chain with random sign disorder}

\begin{figure}
    \centering
    \includegraphics[width=0.9\linewidth]{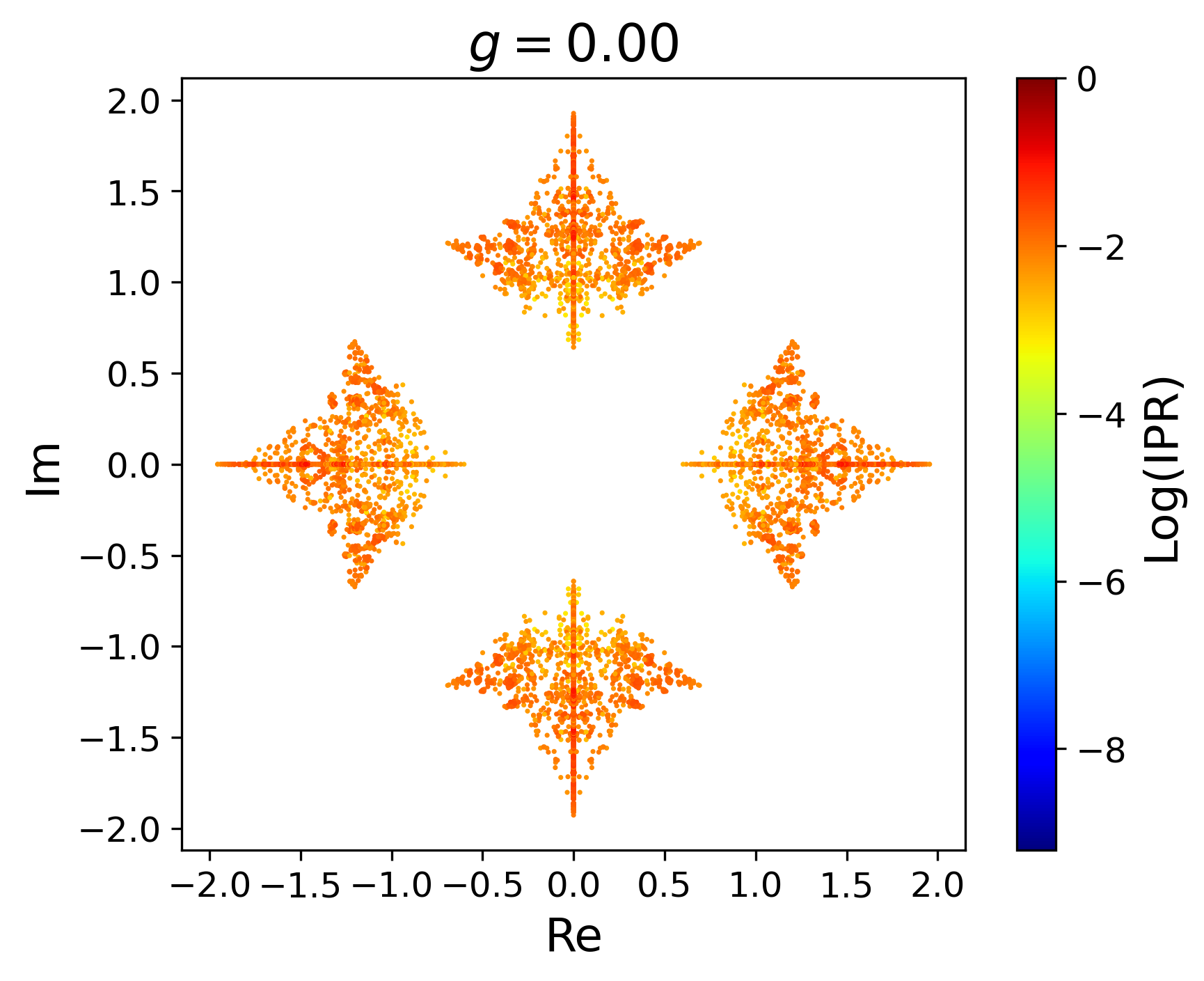}
    \includegraphics[width=0.9\linewidth]{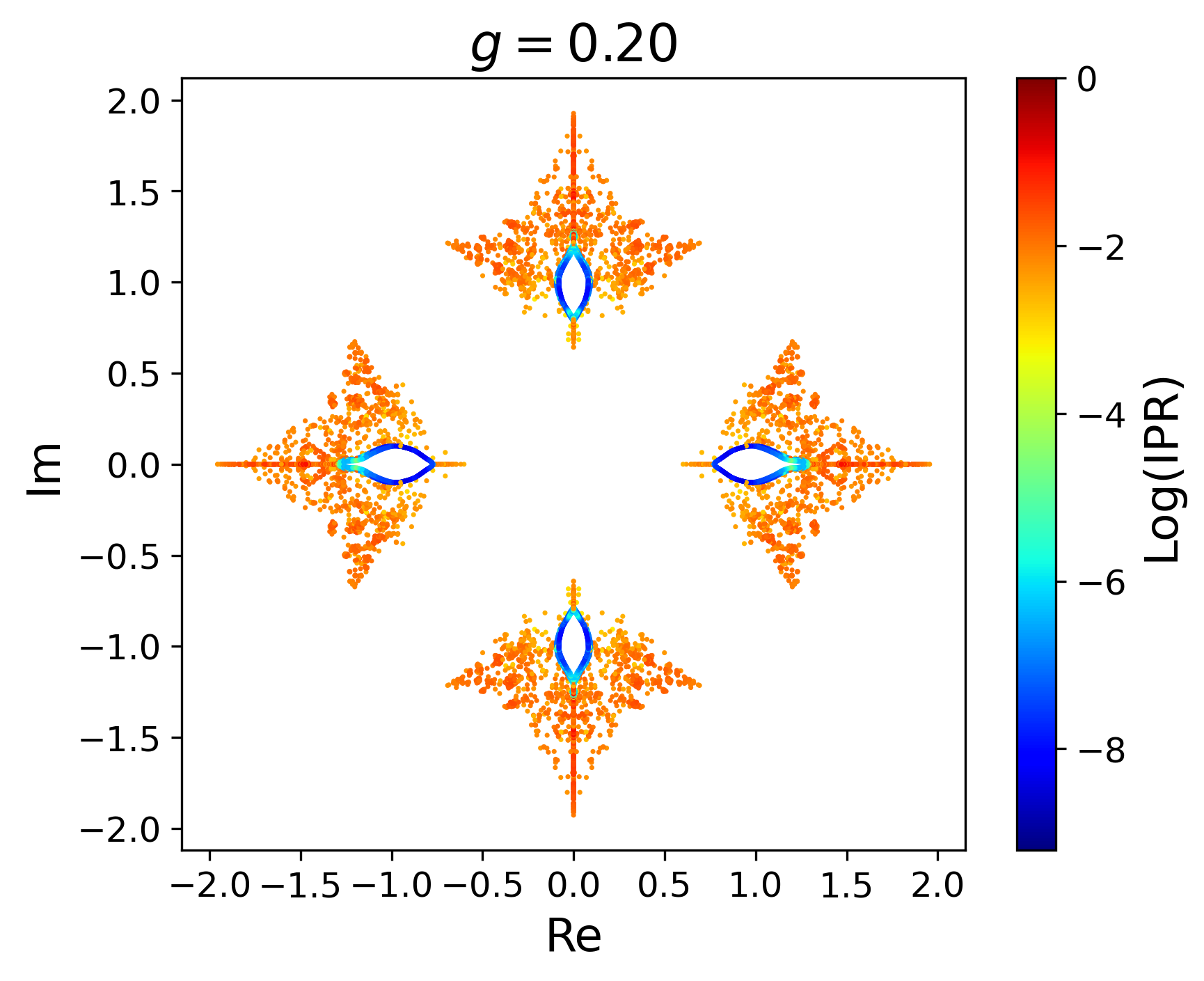}
    \includegraphics[width=0.9\linewidth]{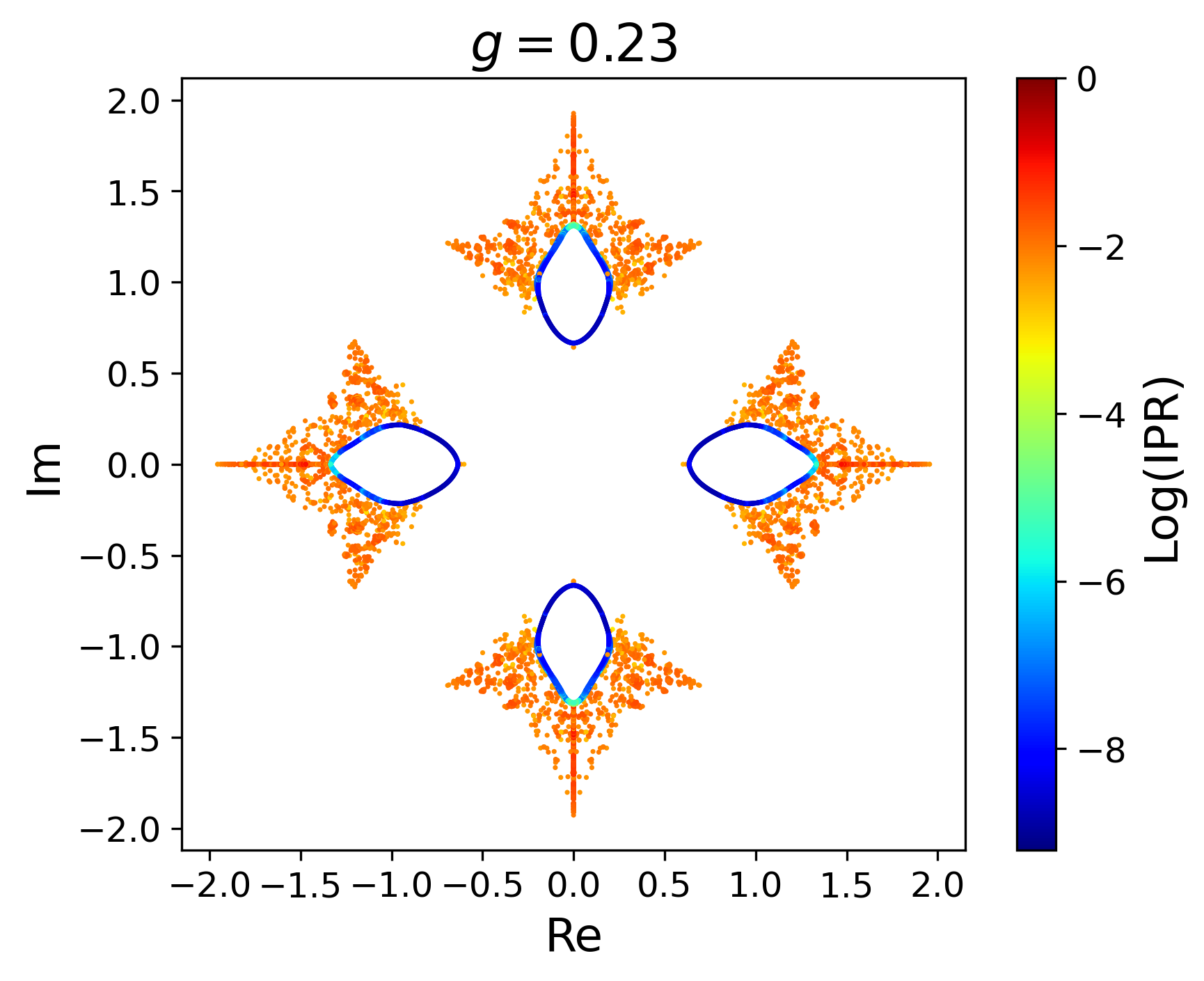}
    \caption{Eigenspectra for the SSH chain with directional bias $g$ and random sign disorder, given by Eq. (\ref{eq:rs_ssh}) with $N=5000$ unit cells and $t=1, \delta t=0.25$. The top plot corresponds to the case of random signs with no directional bias ($g=0$), the middle plot has directional bias $g=0.2$, and the bottom plot has $g=0.23$. The eigenvalues are color-coded according to the inverse participation ratio of their eigenstates, given by Eq. (\ref{eq:pIPR}). Initially, all states are localized, but the directional bias $g$ delocalizes the states, starting from within each band, leading to the four blue loops of extended states.}
    \label{fig:ssh_rs_spectra_1}
\end{figure}

\begin{figure}
    \centering
    \includegraphics[width=0.9\linewidth]{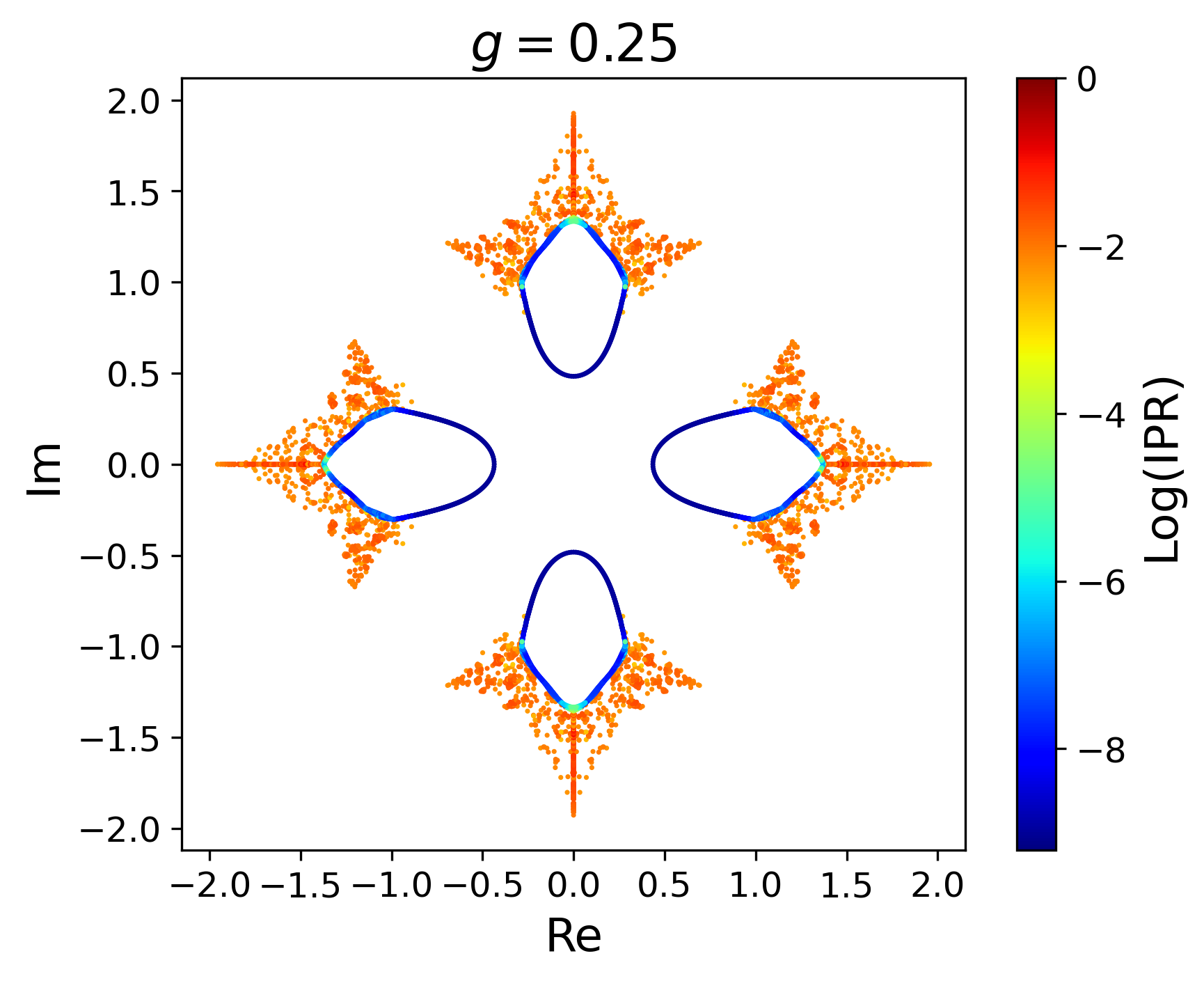}
    \includegraphics[width=0.9\linewidth]{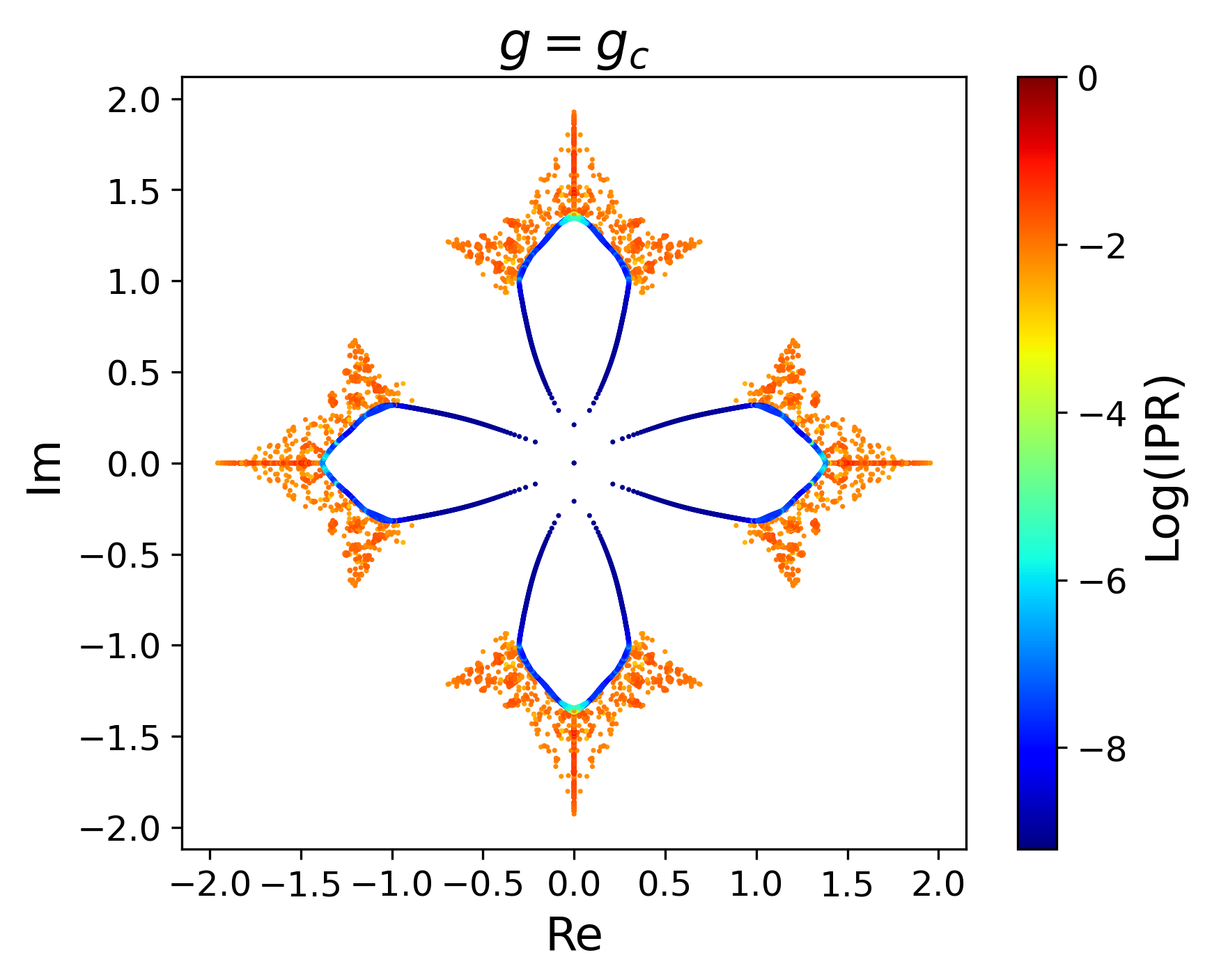}
    \includegraphics[width=0.9\linewidth]{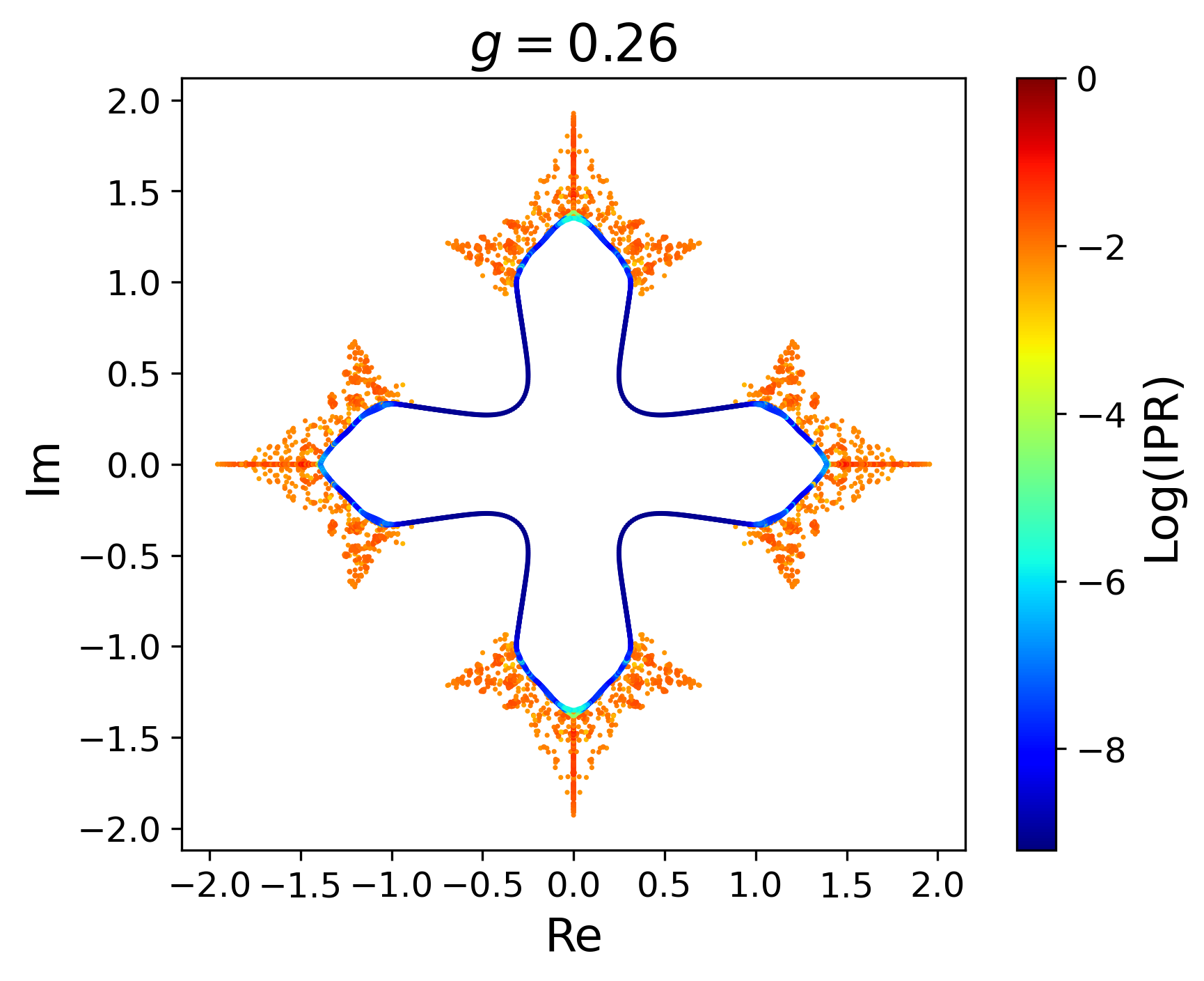}
    \caption{More eigenspectra for the SSH chain with directional bias $g$ and random sign disorder, given by Eq. (\ref{eq:rs_ssh}) with $N=5000$ unit cells and $t=1, \delta t=0.25$. Compared to Figure \ref{fig:ssh_rs_spectra_1}, the directional bias $g$ is increased to larger values, which leads the four blue loops of extended states to expand. Upon increasing $g$ to the exceptional value of $g=g_c\approx0.255$ from Eq. (\ref{eq:ssh_gc}), the four loops of extended states merge, with an exceptional point of order $2$ at the origin. Increasing $g$ further will eventually cause all states to delocalize.}
    \label{fig:ssh_rs_spectra_2}
\end{figure}

Introducing the SSH chain with random sign disorder consistent with Dale's Law in Eq. (\ref{eq:rs_ssh}) leads to striking changes in the eigenspectra in Figures \ref{fig:ssh_rs_spectra_1} and \ref{fig:ssh_rs_spectra_2}, with fractal-like structures reminiscent of \cite{feinberg1999non,amir2016non,zhang2019eigenvalue}. Nevertheless, features of the eigenspectrum for the pure non-Hermitian SSH chain, such as gaps, symmetries, and exceptional points, persist even in the presence of disorder and put constraints on the random eigenspectrum.

\begin{figure}
    \centering
    \includegraphics[width=0.9\linewidth]{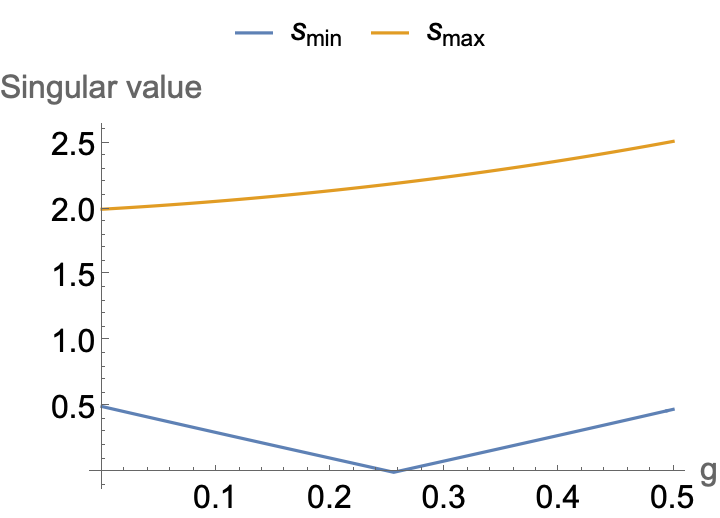}
    \caption{The maximum and minimum singular values (as a function of $g$) that bound the eigenvalue moduli for the non-Hermitian SSH chain ($t=1$, $\delta t=0.25$) with random sign disorder following Dale's law. All eigenvalues $\lambda$ reside in the annulus $s_{\rm min}\le |\lambda|\le s_{\rm max}$ in the complex plane.} 
    \label{fig:ssh_svals}
\end{figure}

In Figures \ref{fig:ssh_rs_spectra_1} and \ref{fig:ssh_rs_spectra_2}, there are regions where the density of states is zero. By computing the singular values of the disordered connectivity matrix $M_S$ in Eq. (\ref{eq:rs_ssh}), we can obtain bounds on the eigenspectrum in the complex plane. Note that $M_S$ can be expressed as the matrix product of the pure model $H_S$ in Eq. (\ref{eq:pure_ssh}) and a diagonal matrix of random signs $\Sigma=\text{diag}(\sigma_{1,A}, \sigma_{1,B}, \cdots, \sigma_{N,A}, \sigma_{N,B})$:
\begin{equation}
    M_S=H_S\cdot\Sigma.
\end{equation}
This relationship is simply a consequence of enforcing Dale's law. Since $\Sigma$ is a diagonal matrix with entries drawn from $\{\pm 1\}$, it is unitary: $\Sigma^\dagger=\Sigma^{-1}$. Thus, the singular values $\{s_i\}$ of the pure model $H_S$ and the disordered model $M_S$ are equal:
\begin{equation}
    \begin{split}
        s_i(M_S)&=\sqrt{\lambda_i(M_S^\dagger M_S)}=\sqrt{\lambda_i(\Sigma^\dagger H_S^\dagger H_S\Sigma)} \\ &=\sqrt{\lambda_i(\Sigma^{-1} H_S^\dagger H_S\Sigma)}=s_i(H_S),
    \end{split}
\end{equation}
where $\{\lambda_i\}$ are the eigenvalues of a matrix. Finding the maximum and minimum singular values is particularly useful because they provide bounds on the norms of eigenvalues: $s_{\rm min} \le |\lambda| \le s_{\rm max}$. Now to compute the singular values, $H_S^\dagger H_S$ is a sparse Hermitian matrix with no disorder:
\begin{equation}
\begin{split}
    H_S^\dagger H_S = \sum_j \Big\{[(t^+e^{-g})^2 & +(t^-e^g)^2]a_j^\dagger a_j \\
     & + (t^+t^-)(a_{j+1}^\dagger a_j+a_j^\dagger a_{j+1}) \Big\} \\
    +\sum_j\Big\{[(t^+e^g)^2 & +(t^-e^{-g})^2]b_j^\dagger b_j \\
    & + (t^+t^-)(b_{j+1}^\dagger b_j+b_j^\dagger b_{j+1}) \Big\},
\end{split}
\end{equation}
and this matrix can be partially diagonalized in Fourier space, giving singular values
\begin{equation}
    \begin{split}
        s_1(k) &= \sqrt{(t^+e^{-g})^2+(t^-e^g)^2+2t^+t^-\cos k}, \\
        s_2(k) &= \sqrt{(t^+e^g)^2+(t^-e^{-g})^2+2t^+t^-\cos k}.
    \end{split}
\end{equation}
This means the minimum and maximum singular values are
\begin{equation} \label{eq:ssh_smin}
    \begin{split}
        s_{\rm min} &= \sqrt{(t^+e^{-g})^2+(t^-e^g)^2-2t^+t^-} \\ 
        & =\sqrt{2t^+t^-[\cosh(2|g-g_c|)-1]},
    \end{split}
\end{equation}
\begin{equation} \label{eq:ssh_smax}
    s_{\rm max} = \sqrt{(t^+e^g)^2+(t^-e^{-g})^2+2t^+t^-},
\end{equation}
as depicted in Figure \ref{fig:ssh_svals}, and the eigenvalues are constrained in absolute value to lie within the annulus $s_{\rm min} \le |\lambda| \le s_{\rm max}$. Note that for the exceptional value $g=g_c$, the smallest singular value is $s_{\rm min}=0$, and so these bounds allow the disordered matrix $M_S$ to have zero modes. 

In fact, provided the number of unit cells $N$ is even, $M_S$ has an exceptional point at $g=g_c$ that survives disorder: there is always a two-fold degenerate $\lambda=0$ eigenvalue, but only a single eigenvector. The left and right eigenvectors (satisfying $M_s^\dagger\mathbf{v}_l=\lambda^*\mathbf{v}_l$ and $M_s\mathbf{v}_r=\lambda\mathbf{v}_r$) are
\begin{equation}
    \mathbf{v}_l = \begin{pmatrix}
        0 \\ 1 \\ 0 \\ -1 \\ \vdots
    \end{pmatrix}\propto \ket{\pi,B}, \;
    \mathbf{v}_r = \begin{pmatrix}
        \sigma_{1,A} \\ 0 \\ -\sigma_{2,A} \\ 0 \\ \vdots
    \end{pmatrix}\propto \Sigma\ket{\pi,A}.
\end{equation}
Since $\mathbf{v}_l$ resides on the $B$ sublattice while $\mathbf{v}_r$ resides on the $A$ sublattice, the state is self-orthogonal ($\mathbf{v}_l^\dagger\mathbf{v}_r=0$), characteristic of an exceptional point. The left eigenvector, which takes the form of a Bloch wave, is identical to that of the model \textit{without} disorder. The right eigenvector is similar, up to the random signs on the $A$ sites.

In addition to bounds on the norms of eigenvalues, there are also symmetries in the spectrum. The same symmetries were also found for the one band, random sign model in \cite{amir2016non}. Because the random connectivity matrix $M_S$ is real, $M_S^*=M_S$ and the spectrum of eigenvalues is symmetric under a reflection about the real axis. In addition, the system also has a chiral symmetry embodied in the relation $\Gamma^{-1}M_S\Gamma=-M_S$, where $\Gamma$ is the diagonal matrix $\Gamma = {\rm diag}(1, -1, 1, -1, \cdots, 1, -1)$. Aside from these two exact symmetries, there is also an approximate statistical symmetry for $\pi/2$ rotations in the complex plane: $M_S\to iM_S$. If we take the diagonal matrix $\Lambda = {\rm diag}(1, \sigma_{1,A}, 1, \sigma_{2,A}, \cdots, 1, \sigma_{N,A})$, the transformed matrix $\Lambda^{-1}M_S\Lambda$ comprises of entries that are functions of products of two random signs $\sigma_i\sigma_j$. Upon sending $M_S\to iM_S$, the products of signs go from $\sigma_i\sigma_j\to -\sigma_i\sigma_j$. Because $\sigma_i\sigma_j$ and $-\sigma_i\sigma_j$ have the same probability distributions, $M_S$ and $iM_S$ have statistically the same spectra. 

Finally, we will also be interested in the localization properties of the eigenstates. Given an eigenstate $\psi$, for Hermitian problems one typically calculates the inverse participation ratio (IPR)
\begin{equation}
    {\rm IPR} = \frac{\sum_i|\psi(i)|^4}{\left(\sum_i|\psi(i)|^2\right)^2},
\end{equation}
which scales as the inverse of the localization length $\xi^{-1}$. But for non-Hermitian systems with distinct left and right eigenvectors $\psi_L$ and $\psi_R$, it is better to consider a gauge-invariant generalization of the inverse participation ratio that depends on the product of the left and right eigenvectors \cite{hatano1998functions, zhang2019eigenvalue}:
\begin{equation} \label{eq:pIPR}
    {\rm IPR}=\frac{\sum_i|\psi_L^*(i)\psi_R(i)|^2}{\left(\sum_i|\psi_L^*(i)\psi_R(i)|\right)^2}.
\end{equation}
In Figures \ref{fig:ssh_rs_spectra_1} and \ref{fig:ssh_rs_spectra_2}, the eigenvalues are color-coded according to their inverse participation ratio. Initially, when there is no directional bias ($g=0$), all states are localized, and the localization length does not diverge anywhere. This is in contrast to one banded models that have disorder only in the off-diagonal hoppings, which leads to a diverging localization at the origin \cite{theodorou1976extended,eggarter1978singular,ziman1982localization,amir2016non}. The localization length remaining finite everywhere is a consequence of enforcing a band gap when setting the parameters $t=1, \delta t=0.25$, and a non-diverging localization length is a feature similar to one band models with on-site disorder, corresponding to disordered self-interactions in neural networks \cite{zhang2019eigenvalue}.

Once we increase the directional bias $g$ to a sufficient level, states begin to delocalize into four loops of extended states, opening up a hole within each band. These loops of extended states expand with increasing $g$, eventually merging at the special value of $g=g_c$ in Eq. (\ref{eq:ssh_gc}) for which the pure non-Hermitian SSH chain has an exceptional point at the origin. As we saw above, this exceptional point still exists for the disordered SSH chain, marking the merging of extended state rings. For large enough $g$, eventually all states will delocalize. 

\subsection{Transfer matrix method for the SSH chain}

In addition to diagonalizing random matrices and computing the inverse participation ratio of the eigenstates, another way of obtaining localization properties for disordered systems involves recasting the matrix equation as a system of transfer matrices or recursion relations and finding the associated Lyapunov exponent \cite{dyson1953dynamics, schmidt1957disordered, derrida2000lyapunov, amir2016non}.

Thus far, we have been solving the eigenvalue problem $M_S\vec{\psi}=\lambda\vec{\psi}$ in Eq. (\ref{eq:rs_ssh}). Away from the boundaries, we can recast this sparse matrix eigenvalue problem as a system of recursive equations:
\begin{equation} \label{eq:recursion_ssh_1}
        t^-e^{-g}\sigma_{j-1,B}\psi_{j-1,B} + t^+e^g\sigma_{j,B}\psi_{j,B} = \lambda \psi_{j,A},
\end{equation}
\begin{equation} \label{eq:recursion_ssh_2}
    t^+e^{-g}\sigma_{j,A}\psi_{j,A} + t^-e^g\sigma_{j+1,A}\psi_{j+1,A}=\lambda \psi_{j,B}.
\end{equation}
A wavefunction with localization length $\xi$ and a center of localization, say, at the origin, takes the form $|\psi(x)|\sim e^{-x/\xi}$. To estimate the localization length using the recursion relations, it is useful to recast Eqs. (\ref{eq:recursion_ssh_1}) and (\ref{eq:recursion_ssh_2}) in terms of the Ricatti variables $r_j^+=\psi_{j,B}/\psi_{j,A}$ and $r_j^-=\psi_{j,A}/\psi_{j-1,B}$:
\begin{equation} \label{eq:recursion_ssh}
    \begin{split}
        r_j^+ &=\frac{1}{t^+e^g\sigma_{j,B}}\left[\lambda-\frac{t^-e^{-g}\sigma_{j-1,B}}{r_j^-}\right], \\
        r_{j+1}^- &=\frac{1}{t^-e^g\sigma_{j+1,A}}\left[\lambda-\frac{t^+e^{-g}\sigma_{j,A}}{r_j^+}\right].
    \end{split}
\end{equation}
The Lyapunov exponent $\gamma$, which is equal to the inverse of the localization length $\xi^{-1}$, is related to the Ricatti variables through
\begin{equation} \label{eq:ssh_lyapunov_formula}
    \gamma(\lambda)=\lim_{N\to\infty}\frac{1}{2N}\sum_{j=1}^N \log |r_j^+r_{j+1}^-|.
\end{equation}
Compared to directly diagonalizing matrices, this method provides a faster way to estimate the localization of a state corresponding to an eigenvalue $\lambda$. However, a downside is that this method always returns an estimate for $\gamma(\lambda)$, regardless of whether or not there actually exists an eigenstate for $\lambda$. 

In addition, the recursion relations in Eq. (\ref{eq:recursion_ssh}) provide a connection between the directional biasing parameter $g$ and the delocalization of all eigenstates. Upon making a change of variables
\begin{equation}
    \rho_j^+ = e^gr_j^+, \quad \rho_j^- = e^gr_j^-,
\end{equation}
note that we recover the $g=0$ recursion relations:
\begin{equation}
    \begin{split}
        \rho_j^+ &=\frac{1}{t^+\sigma_{j,B}}\left[\lambda - \frac{t^-\sigma_{j-1,B}}{\rho_j^-}\right], \\
        \rho_{j+1}^- &=\frac{1}{t^-\sigma_{j+1,A}}\left[\lambda - \frac{t^+\sigma_{j,A}}{\rho_j^+}\right].
    \end{split}
\end{equation}
Thus, the effect of increasing $g$ is to lower the Lyapunov exponent $\gamma$ in Eq. (\ref{eq:ssh_lyapunov_formula}) by $g$. The criterion for the delocalization of a particular eigenstate that has localization length $\xi$ when $g=0$ is given by $g=\gamma=\xi^{-1}$. In particular, the Lyapunov exponent at the origin ($\lambda=0$) is determined analytically to be
\begin{equation}
    \gamma(\lambda=0)=g_c-g,
\end{equation}
which is consistent with our earlier observation that the merging of the rings of extended states in Figure \ref{fig:ssh_rs_spectra_2} occurred at the origin for $g=g_c$.

\begin{figure}
    \centering
    \includegraphics[width=0.9\linewidth]{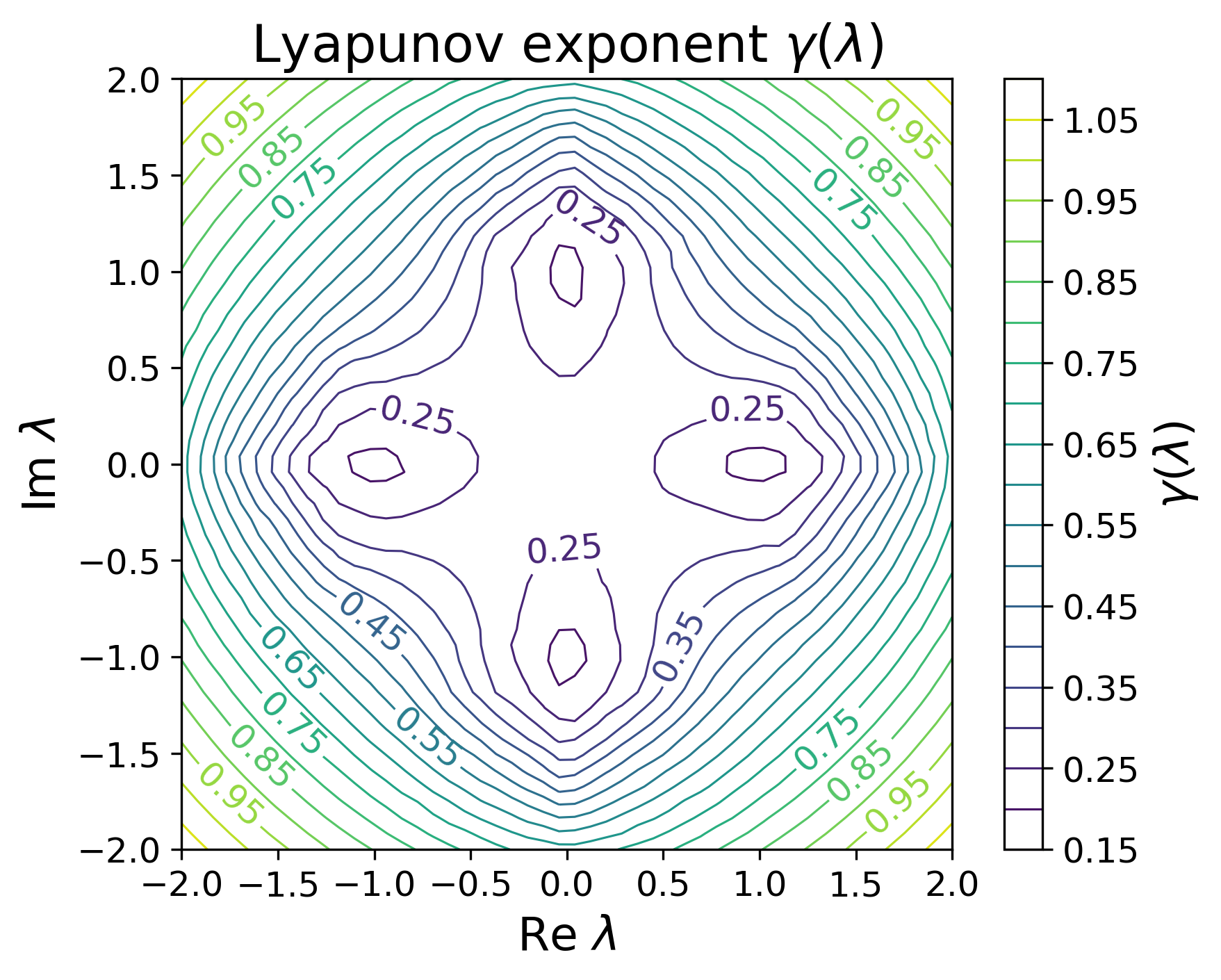}
    \includegraphics[width=0.9\linewidth]{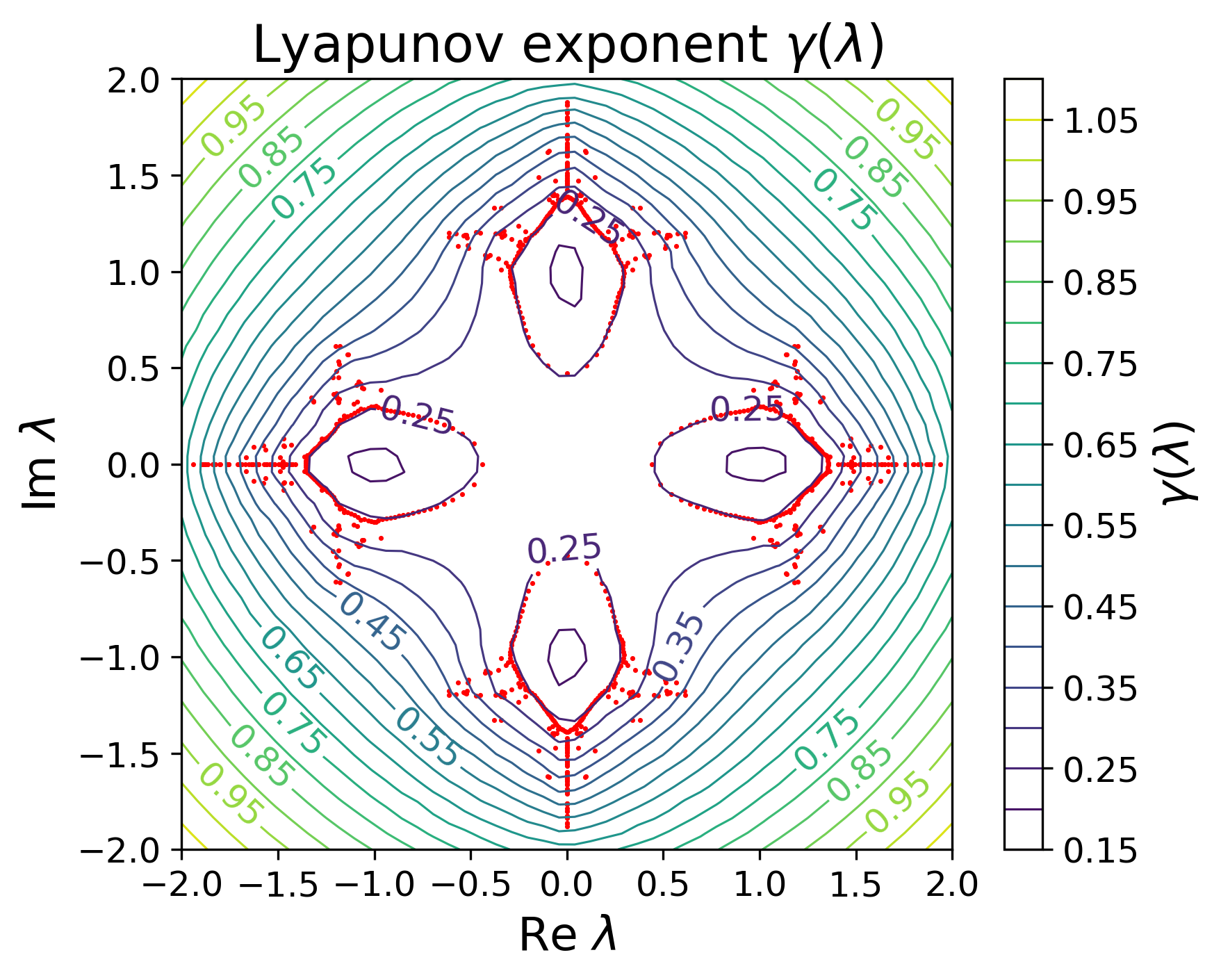}
    \caption{(Top) Contour plot of the Lyapunov exponent $\gamma(\lambda)$ as a function of a complex eigenvalue $\lambda$ for the random sign SSH chain with $t=1, \delta t= 0.25$, and $g=0$. (Bottom) Superimposed on the contours is the eigenspectrum for a particular instance of disorder ($N=500$ unit cells, $t=1, \delta t=0.25$, and $g=0.25$). The loops of extended states reside on the $\gamma=0.25$ contour, in agreement with the delocalization criterion $g=\gamma=\xi^{-1}$.}
    \label{fig:ssh_lyapunov_contours}
\end{figure}

In Figure \ref{fig:ssh_lyapunov_contours}, we used the method described in Eq. (\ref{eq:ssh_lyapunov_formula}) to numerically estimate the localization properties of the random sign SSH chain.
To generate the plots, we used $N = 10000$ steps and dropped the first 10\% of terms to reduce the effects of initial conditions. In the top plot, we evaluated the Lyapunov exponent $\gamma(\lambda)$ for $g=0$ over the complex plane and plotted a selected subset of contours of constant $\gamma$. The minimum value of the Lyapunov exponent over the complex plane is not zero, and it does not occur at the origin, but rather within the bands at values around $\pm 1$ and $\pm i$. In the bottom plot, we have superimposed a sample eigenspectrum for a particular random matrix with $N=500$ unit cells, $t=1, \delta t=0.25$, and $g=0.25$. The rings of extended states for this matrix lies precisely on the $\gamma=0.25$ contour, which is what we expect from the delocalization criterion $g=\gamma=\xi^{-1}$ discussed above. 

\section{The non-Hermitian ladder} \label{sec:nhladder}

\begin{figure}
    \centering
    \includegraphics[width=1\linewidth]{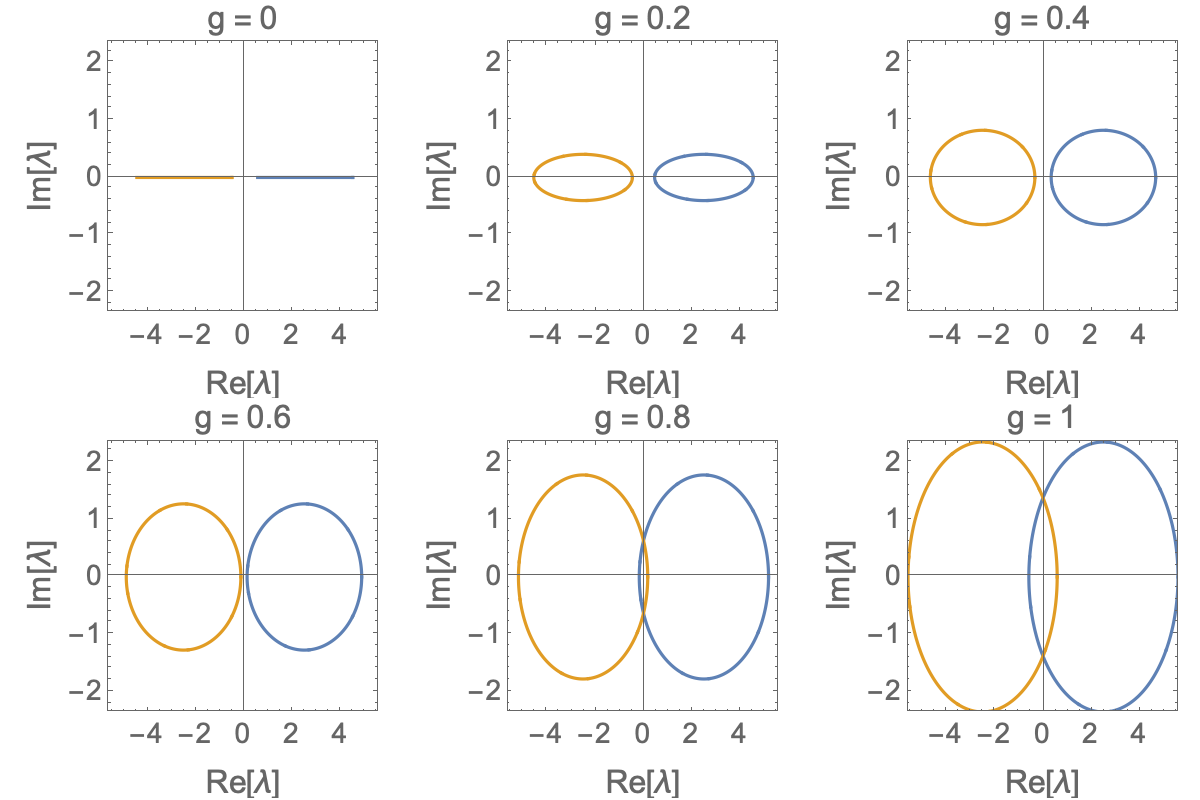}
    \caption{Eigenspectrum for the non-Hermitian ladder for $t=1$ and $u=2.5$. As the non-Hermitian parameter $g$ is increased, eigenvalues leave the real axis. The band gap closes at the diabolic point given by $g_c=\cosh^{-1}(u/2t)\approx 0.693$.} 
    \label{fig:ladder_bands}
\end{figure}

In this section, we will study the spectrum and localization properties of the ladder model with random sign disorder in Eq. (\ref{eq:rs_ladder}), and we will see how its localization properties differ from that of the SSH chain in the previous section. As before, it will be useful to first analyze the pure ladder model $H_L$, as its properties will have consequences even once we include the disorder:
\begin{equation} \label{eq:pure_ladder}
    \begin{split}
        H_L = t &\sum_{j=1}^N (e^g\ket{j, A}\bra{j+1, A} + e^{-g}\ket{j+1, A}\bra{j, A})\\
        & + t\sum_{j=1}^N (e^g\ket{j, B}\bra{j+1, B} + e^{-g}\ket{j+1, B}\bra{j, B})\\
        &+u\sum_{j=1}^N(\ket{j,A}\bra{j,B}+\ket{j,B}\bra{j,A}).
    \end{split}
\end{equation}
As before, upon assuming periodic boundary conditions $\ket{j+N,A}=\ket{j,A}$ and $\ket{j+N,B}=\ket{j,B}$, the system can be partially diagonalized using Bloch eigenfunctions $\ket{k,A}=\frac{1}{\sqrt{N}}\sum_j e^{ikj}\ket{j,A}$ and $\ket{k,B}=\frac{1}{\sqrt{N}}\sum_j e^{ikj}\ket{j,B}$. For each momentum $k$ in the first Brillouin zone, we have the two level system
\begin{equation}
    H_L(k)=\begin{pmatrix}
         t[e^{g+ik}+e^{-(g+ik)}] & u \\
         u & t[e^{g+ik}+e^{-(g+ik)}]
    \end{pmatrix},
\end{equation}
with corresponding complex eigenvalues 
\begin{equation} \label{eq:ladder_bands}
    E_\pm(k) = 2t[\cosh g \cos k + i \sinh g \sin k] \pm u.
\end{equation}
Provided that $u>2t$, the Hermitian ($g=0$) ladder has two distinct bands. When we increase the directional bias parameter $g$ to be nonzero, the bands pop into the complex plane and eventually touch at a single point ($E_-(k=0)=E_+(k=\pi)=0$) for
\begin{equation} \label{eq:ladder_gc}
    g_c=\cosh^{-1}\left(\frac{u}{2t}\right).
\end{equation}
This spectral evolution is illustrated in Figure \ref{fig:ladder_bands}, where the ladder band structure from Eq. (\ref{eq:ladder_bands}) has been plotted for $t=1, u=2.5$, and various values of $g$. 

In contrast to the non-Hermitian SSH chain, the resulting two-fold degenerate zero eigenvalue is not an \textit{exceptional point}, but rather a \textit{diabolic point} \cite{yarkony1996diabolical, ashida2020non}, since it corresponds to two distinct eigenvectors, as follows from considering the $4\times 4$ subspace of Bloch states with $k=0$ and $\pi$: 
\begin{equation}
    \begin{split}
        &\left(\begin{array}{c|c}
        H_L(k=0; g=g_c) & 0 \\
        \hline
        0 & H_L(k=\pi; g=g_c)
    \end{array}\right) \\
    & \hspace{1.5cm}= u\begin{pmatrix}
        1 & 1 & 0 & 0 \\
        1 & 1 & 0 & 0 \\
        0 & 0 & -1 & 1 \\
        0 & 0 & 1 & -1
    \end{pmatrix}.
    \end{split}
\end{equation}
This matrix is Hermitian and has two distinct, orthogonal zero modes $\frac{1}{\sqrt{2}}(\ket{0,A}-\ket{0,B})$ and $\frac{1}{\sqrt{2}}(\ket{\pi,A}+\ket{\pi,B})$, as well as eigenmodes $\frac{1}{\sqrt{2}}(\ket{0,A}+\ket{0,B})$ and $\frac{1}{\sqrt{2}}(\ket{\pi,A}-\ket{\pi,B})$, with eigenvalues $2u$ and $-2u$, respectively. 

\subsection{Ladder with random sign disorder}

\begin{figure}
    \centering 
    \includegraphics[width=0.83\linewidth]{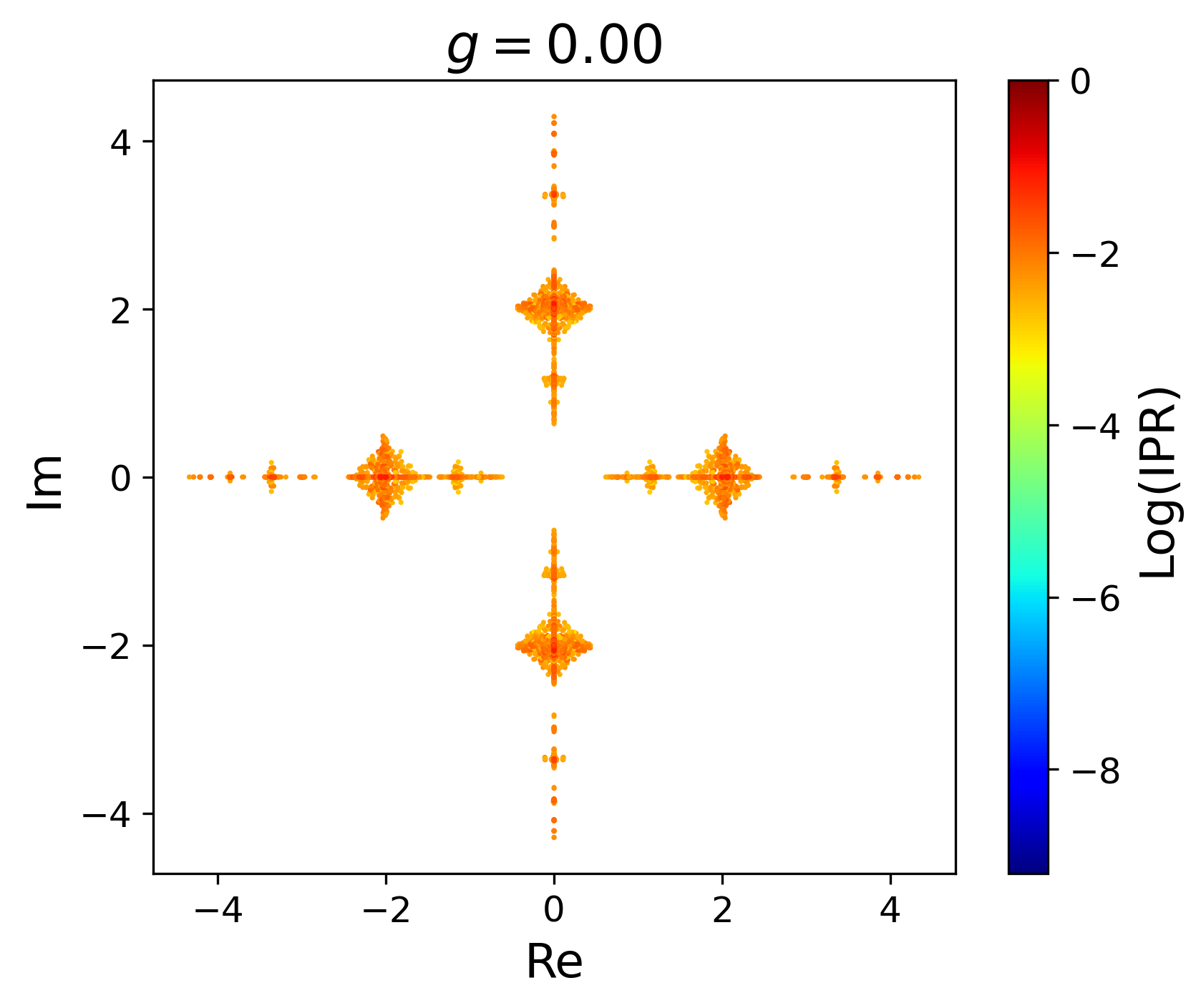}
    \includegraphics[width=0.83\linewidth]{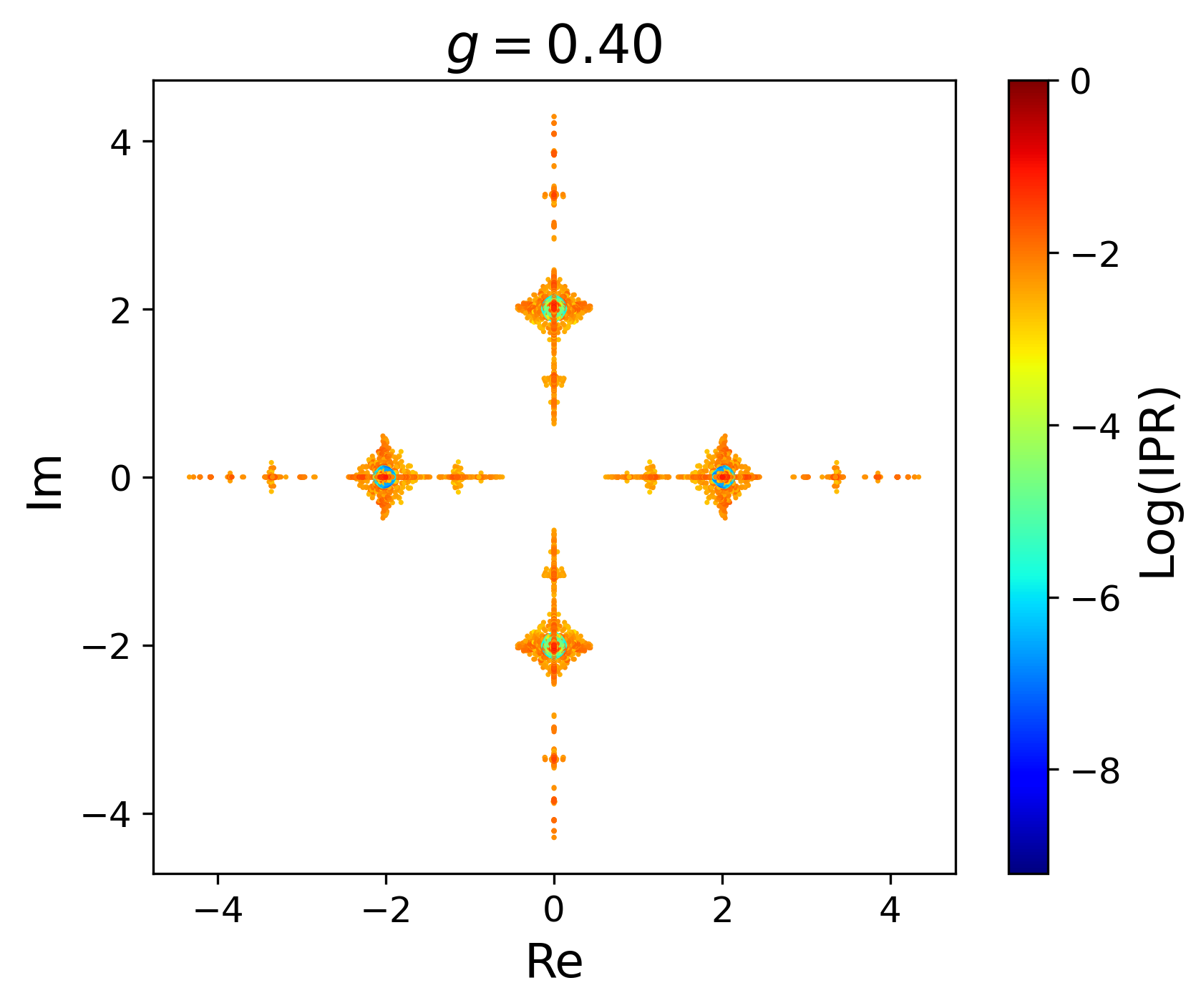}
    \includegraphics[width=0.83\linewidth]{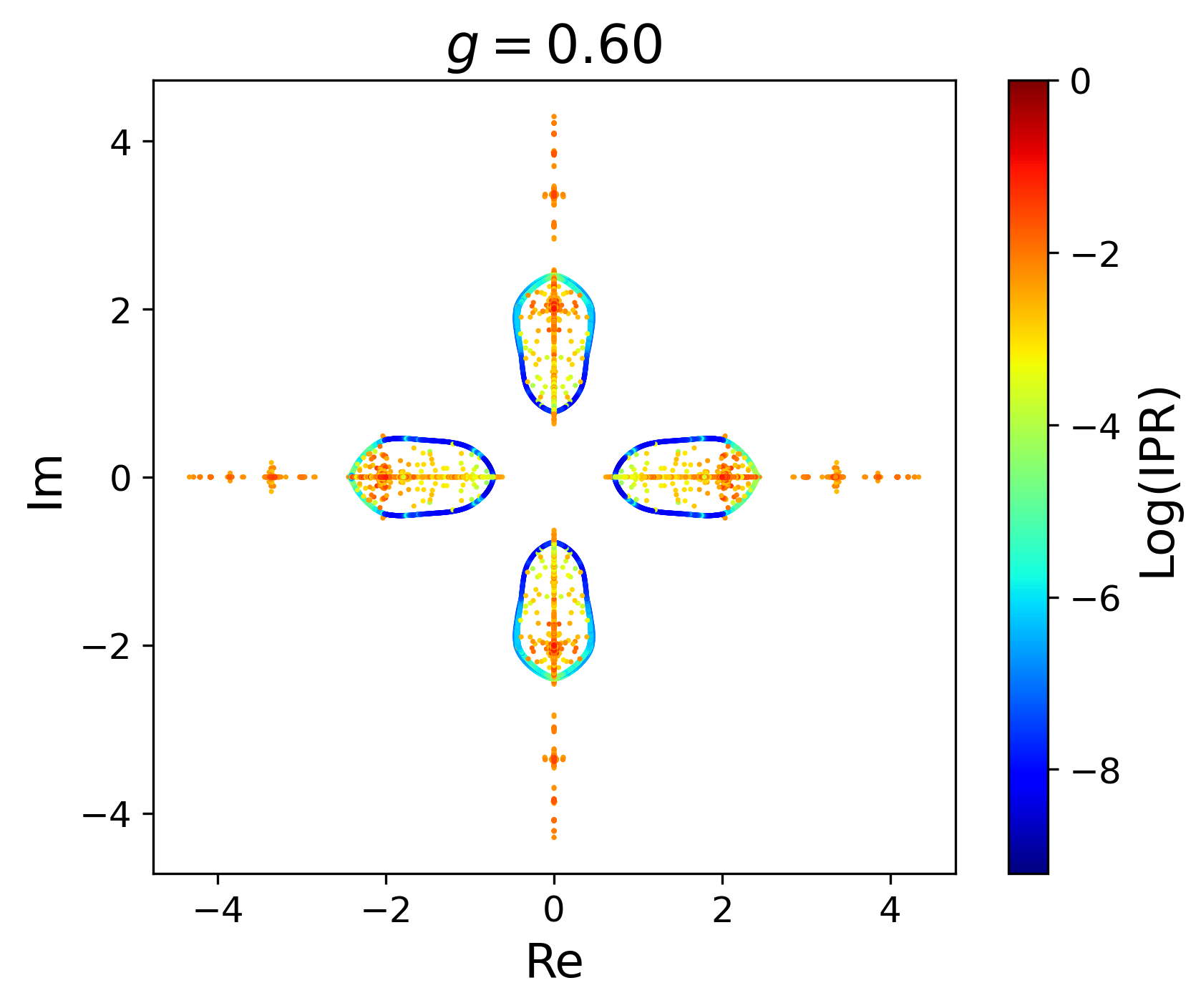}
    \caption{Eigenspectra for the ladder model with directional bias $g$ and random sign disorder, given by Eq. (\ref{eq:rs_ladder}) with $N=5000$ unit cells and $t=1, u=2.5$. The top plot corresponds to the case with no directional bias ($g=0$), the middle plot has $g=0.4$, and the bottom plot has $g=0.6$. The eigenvalues are color-coded according to the inverse participation ratio of their eigenstates, given by Eq. (\ref{eq:pIPR}). The directional bias $g$ delocalizes the states, starting from within each of the four diamond-shaped eigenvalue clusters centered on the real and imaginary axes in the complex plane. In contrast to the SSH chain, when states initially start to delocalize into loops of extended states, they do not open a hole in the spectrum, as there are localized states still inside each loop.}
    \label{fig:ladder_rs_spectra_1}
\end{figure}

\begin{figure}
    \centering
    \includegraphics[width=0.83\linewidth]{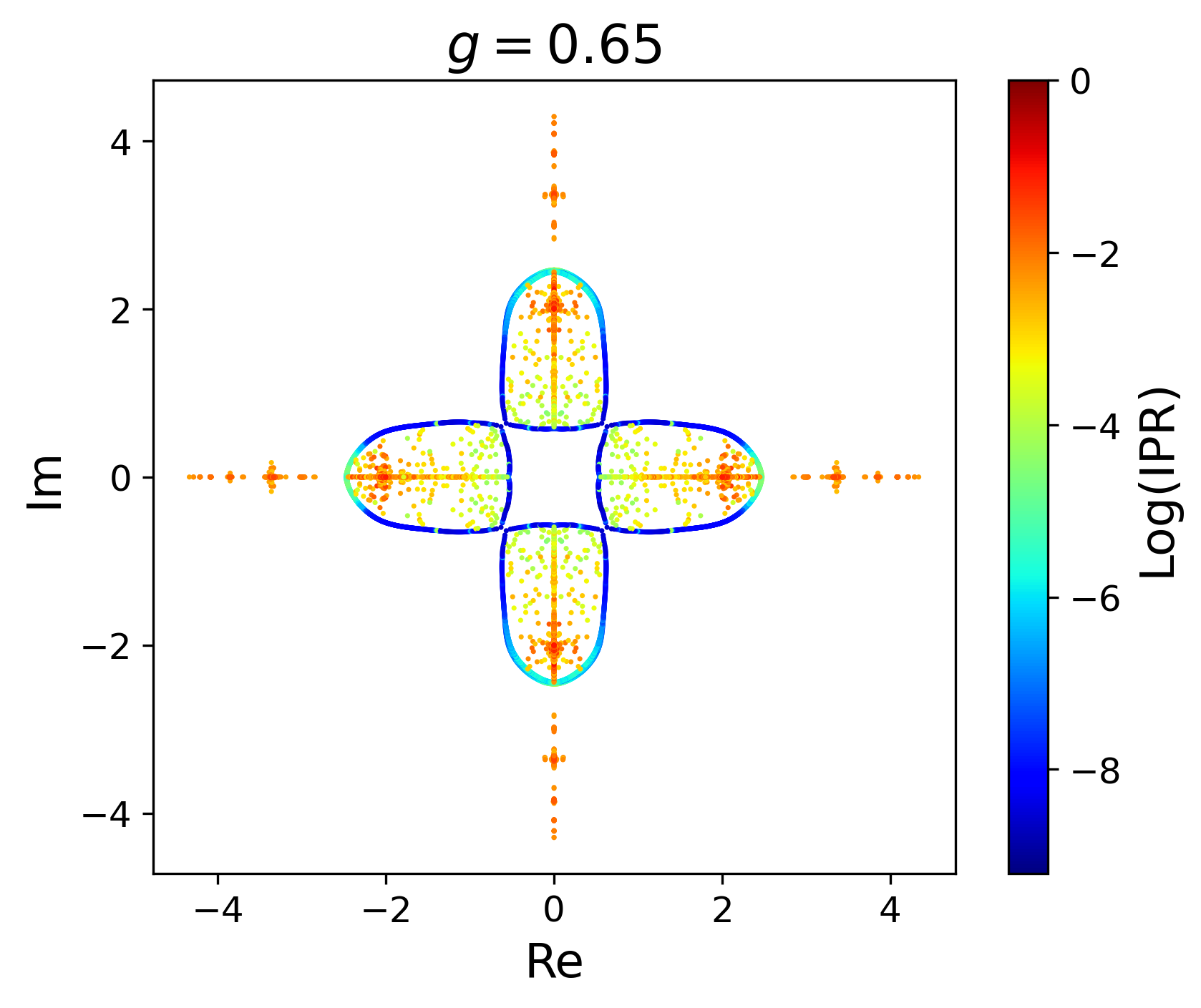}
    \includegraphics[width=0.83\linewidth]{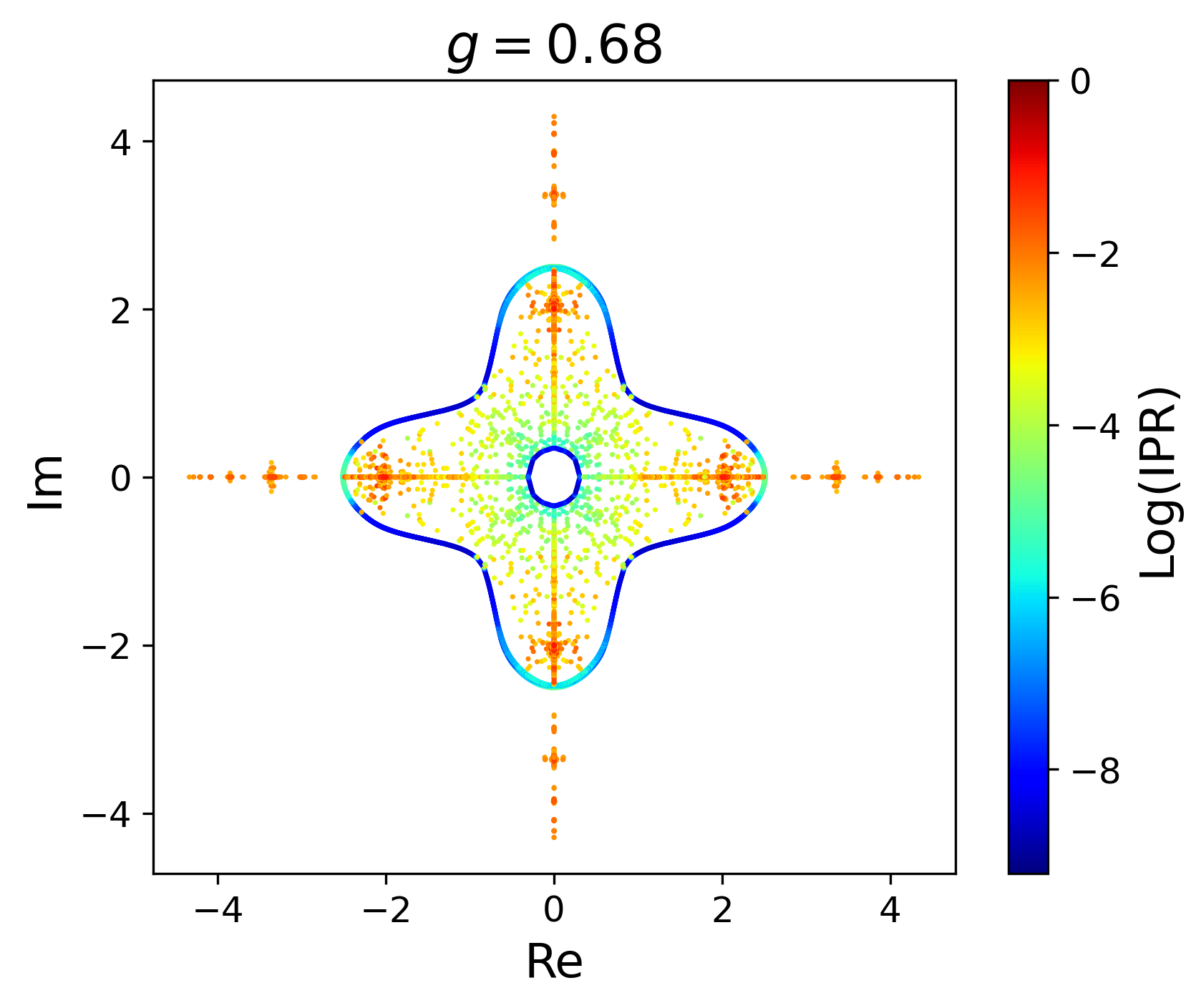}
    \includegraphics[width=0.83\linewidth]{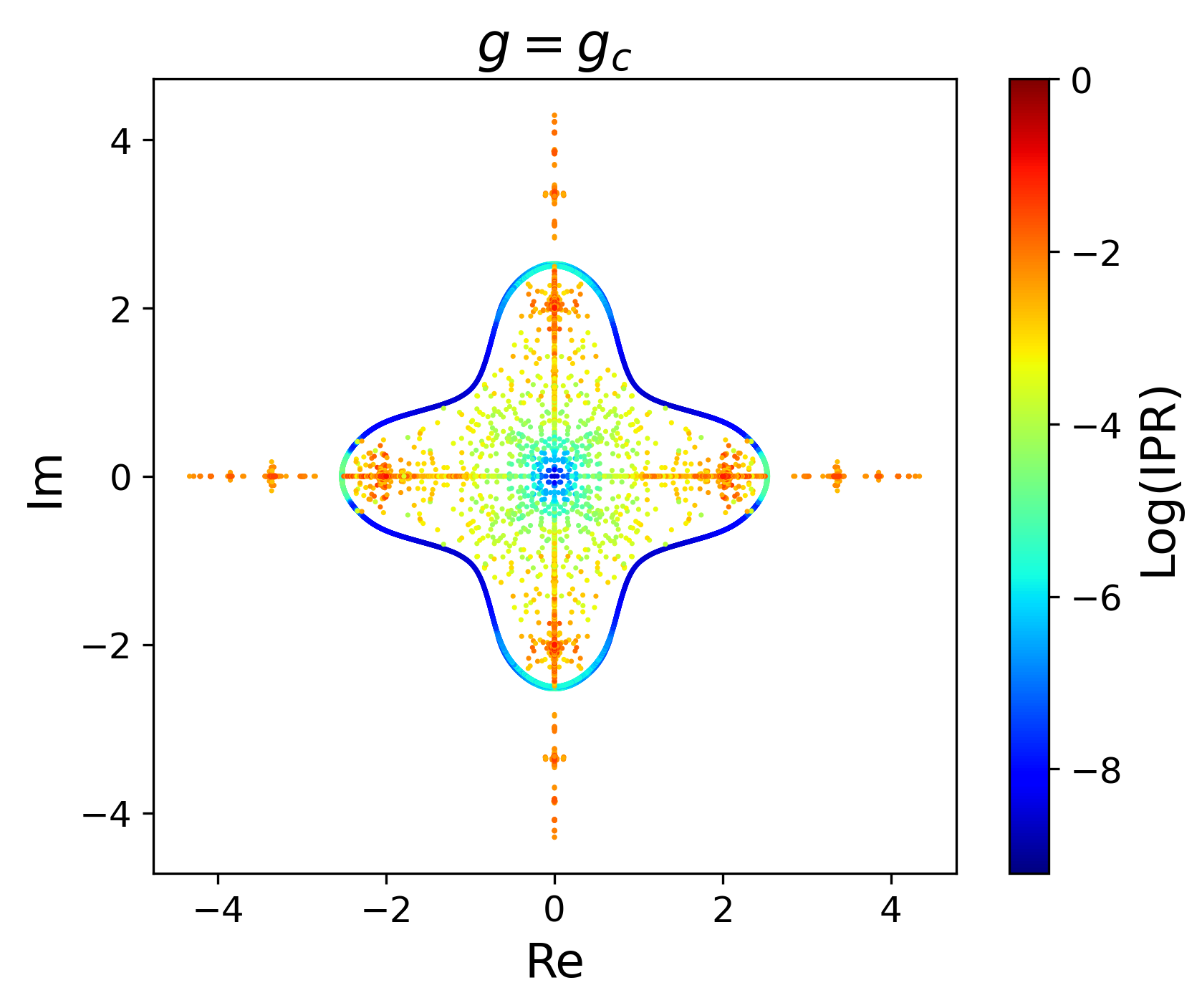}
    \caption{More eigenspectra for the ladder model with directional bias $g$ and random sign disorder, given by Eq. (\ref{eq:rs_ladder}) with $N=5000$ unit cells and $t=1, u=2.5$. Compared to Figure \ref{fig:ladder_rs_spectra_1}, the directional bias $g$ is increased to larger values, which leads the four loops of extended states to further expand, though with some localized states still inside each loop. Eventually the four loops of extended states merge to form two loops of extended states, with localized states in between, as shown in the lower left spectrum. The inner loop shrinks and vanishes upon increasing $g$ to the diabolic value of $g=g_c\approx0.693$ from Eq. (\ref{eq:ladder_gc}).}
    \label{fig:ladder_rs_spectra_2}
\end{figure}

\begin{figure}
    \centering
    \includegraphics[width=0.83\linewidth]{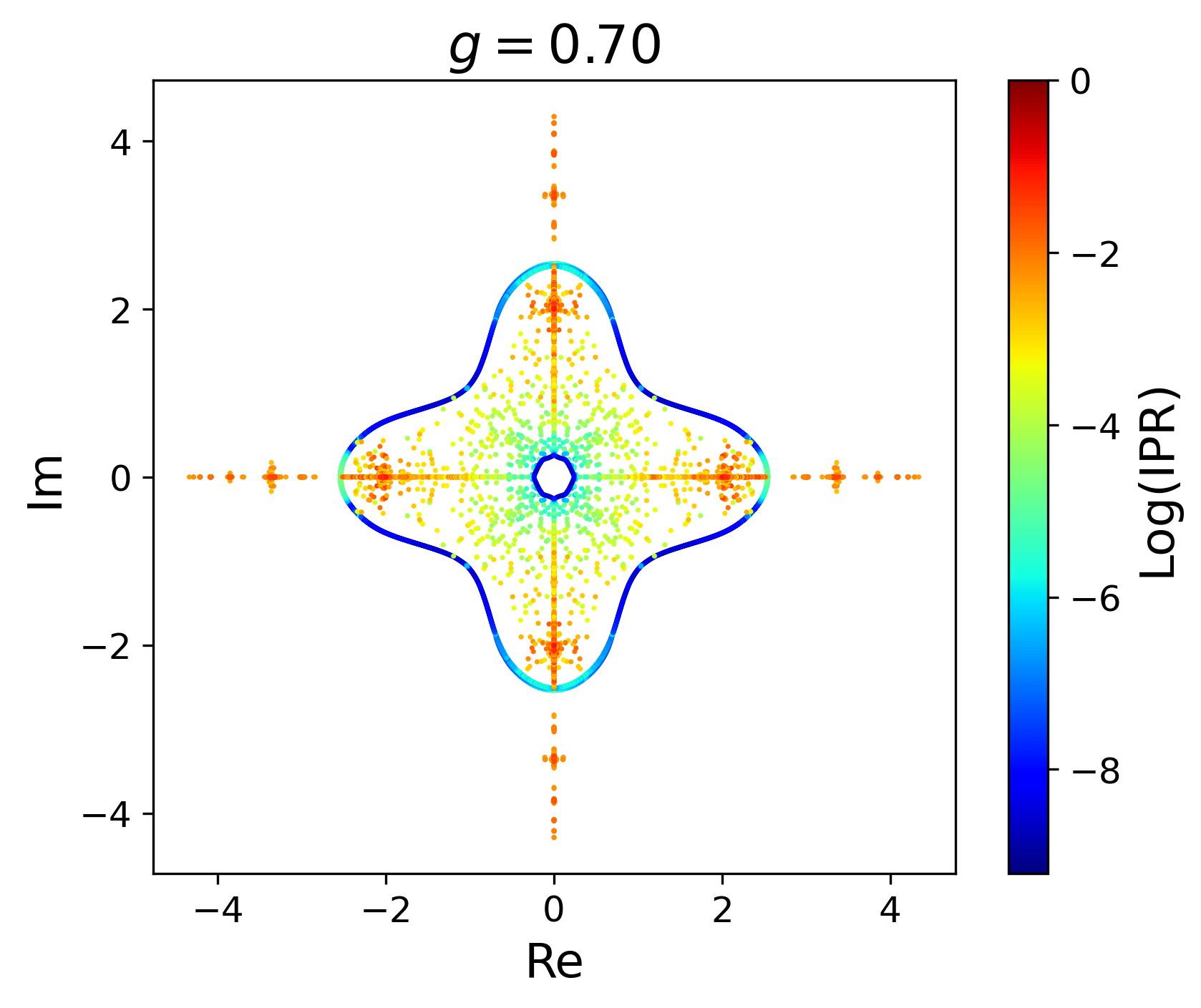}
    \includegraphics[width=0.83\linewidth]{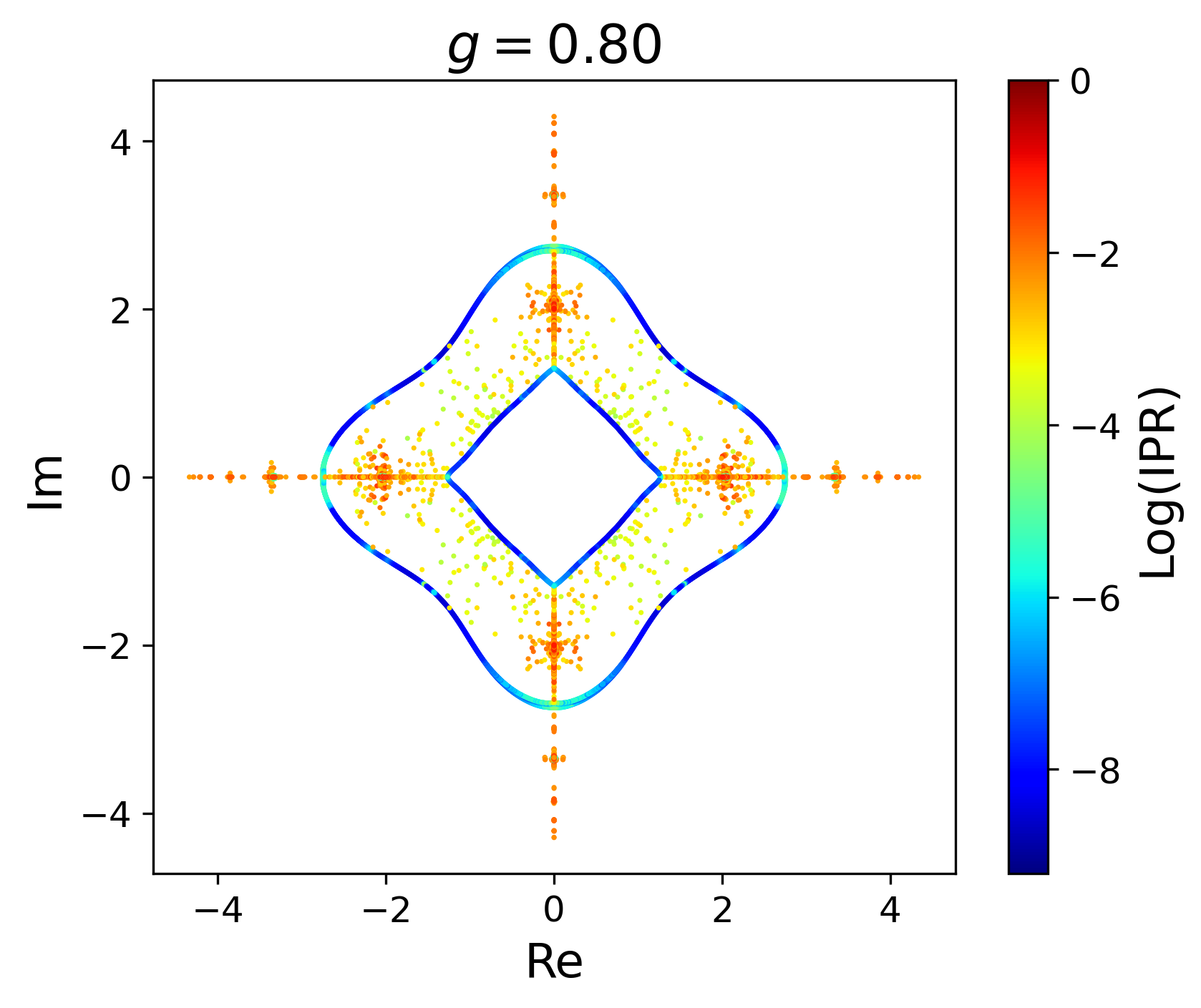}
    \includegraphics[width=0.83\linewidth]{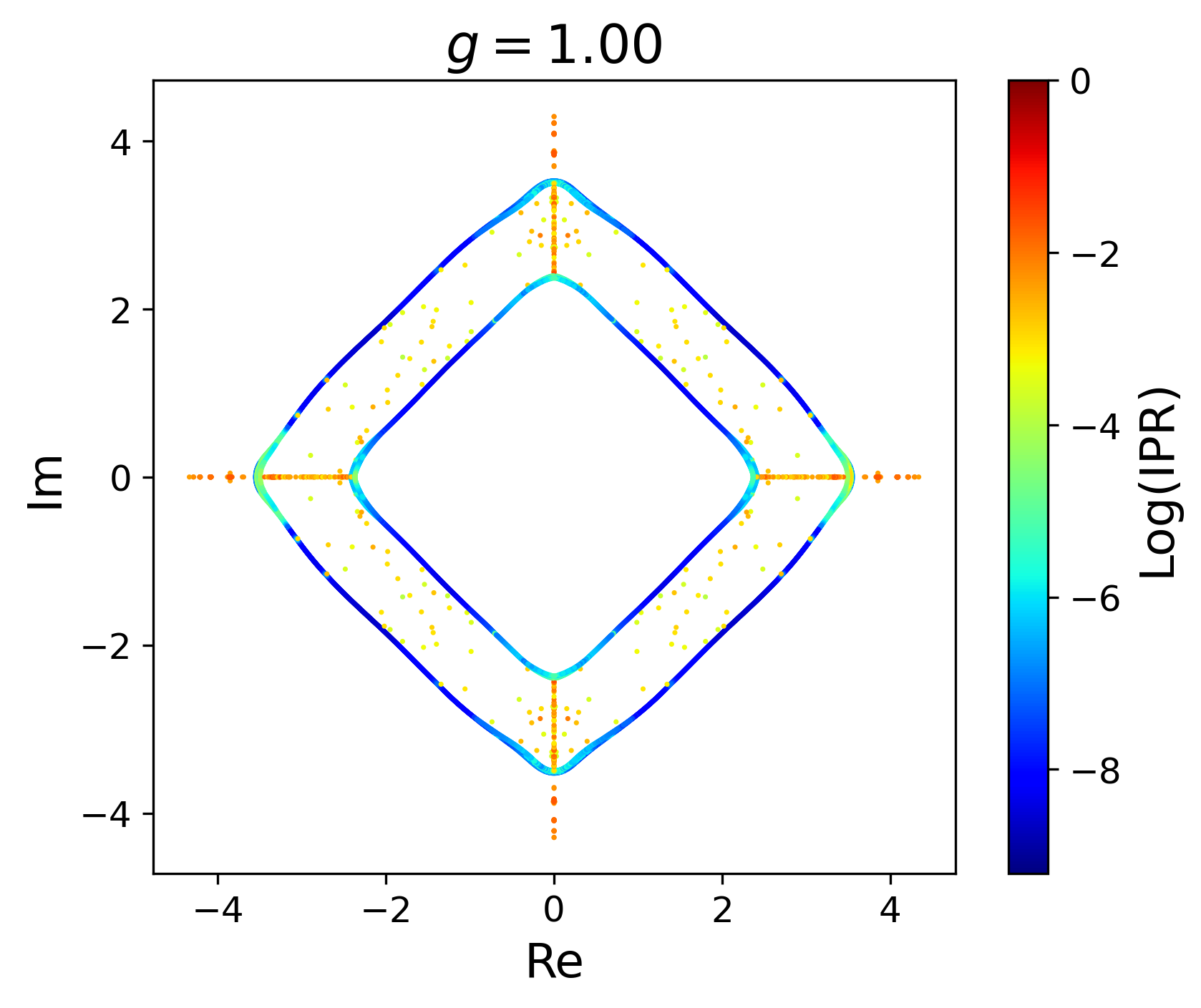}
    \caption{More eigenspectra for the ladder model with directional bias $g$ and random sign disorder, given by Eq. (\ref{eq:rs_ladder}) with $N=5000$ unit cells and $t=1, u=2.5$. Compared to Figures \ref{fig:ladder_rs_spectra_1} and \ref{fig:ladder_rs_spectra_2}, the directional bias $g$ is increased to even larger values, with the directional bias now above its diabolic value $g>g_c\approx 0.693$. In this regime, a new hole in the spectrum opens about the origin, increasing with $g$. Further increasing $g$ will eventually cause all states to delocalize.}
    \label{fig:ladder_rs_spectra_3}
\end{figure}

As before, the random sign disorder in Eq. (\ref{eq:rs_ladder}) leads to a striking change in the eigenspectrum, as shown in Figures \ref{fig:ladder_rs_spectra_1}, \ref{fig:ladder_rs_spectra_2}, and \ref{fig:ladder_rs_spectra_3}. Before discussing the delocalization of states that occurs upon increasing the directional bias $g$, we summarize some of the symmetries and constraints on the random eigenspectrum, using the same methods outlined for the SSH chain in the previous section.

\begin{figure}
    \centering
    \includegraphics[width=0.9\linewidth]{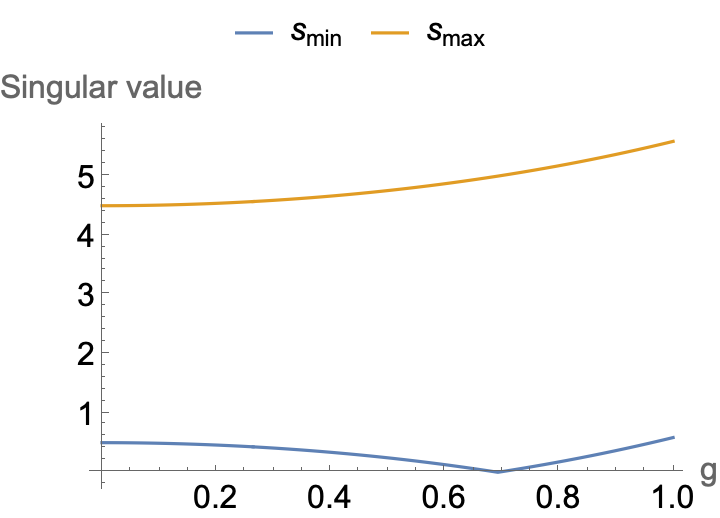}
    \caption{The maximum and minimum singular values (as a function of $g$) for the non-Hermitian ladder ($t=1$, $u=2.5$) with random sign disorder following Dale's law. All eigenvalues $\lambda$ reside in the annulus $s_{\rm min}\le |\lambda|\le s_{\rm max}$ in the complex plane.}
    \label{fig:ladder_svals}
\end{figure}

The eigenvalues are constrained to reside in an annulus in the complex plane, with inner and outer radius given by the smallest and largest singular values. To compute the singular value, we invoke the same argument as for the SSH chain: Dale's law (all neurons are purely inhibitory or excitatory) means the disordered matrix $M_L$ is given by the matrix product
\begin{equation} \label{eq:ladder_disorder_product}
    M_L = H_L\cdot \Sigma,
\end{equation}
where $H_L$ is the ladder model without disorder, and $\Sigma$ is a diagonal matrix with entries $\sigma_{1,A}, \sigma_{1,B}, \cdots, \sigma_{N,A}, \sigma_{N,B}$ on the diagonal. This means the singular values for $M_L$ and $H_L$ are the same. The singular values for $H_L$ are given by the square root of the eigenvalues of $H_L^\dagger H_L$. We find that the sparse Hermitian matrix
\begin{equation}
\begin{split}
    H_L^\dagger H_L = & \sum_j [u^2+2t^2\cosh g](a_j^\dagger a_j+b_j^\dagger b_j) \\
    & + \sum_j 2ut(\cosh g)(a_j^\dagger b_{j+1} + b_j^\dagger a_{j+1} + h.c.) \\
    & + \sum_j t^2(a_{j+2}^\dagger a_j + b_{j+2}^\dagger b_j + h.c.) \\
\end{split}
\end{equation}
can be partially diagonalized in Fourier space, leading to singular values
\begin{equation}
    s_\pm(k)=\sqrt{u^2+2t^2\cosh 2g+2t^2\cos 2k \pm 4ut\cosh g\cos k}.
\end{equation}
In the case where $u>2t$, the minimum and maximum singular values are then
\begin{equation}
        s_{\rm min}=|u-2t\cosh g|, \quad s_{\rm max}=u+2t\cosh g,
\end{equation}
forming bounds which are shown in Figure \ref{fig:ladder_svals}. Note that for the diabolic value $g=g_c$ from Eq. (\ref{eq:ladder_gc}), the smallest singular value is $s_{\rm min}=0$, and so these bounds allow the disordered matrix $M_L$ to have zero modes. 

Furthermore, if the number of unit cells $N$ is even, the disordered ladder $M_L$ still has a diabolic point at $g=g_c$. Since $M_L=H_L\cdot\Sigma$ from Eq. (\ref{eq:ladder_disorder_product}), there is a two-fold degenerate $\lambda=0$ eigenvalue with a complete set of eigenstates, with the space of left and right eigenvectors given by
\begin{equation}
    \begin{split}
        \mathbf{v}_l &\in {\rm span}\left\{\ket{0,A}-\ket{0,B}, \ket{\pi,A}+\ket{\pi,B}\right\}, \\
        \mathbf{v}_r &\in {\rm span}\left\{\Sigma(\ket{0,A}-\ket{0,B}), \Sigma(\ket{\pi,A}+\ket{\pi,B})\right\}.
    \end{split}
\end{equation}

In addition to bounds on the norms of eigenvalues, there are also symmetries in the spectrum, revealed by similar arguments as for the SSH chain and in \cite{amir2016non}. Because the random connectivity matrix $M_L$ is real, there is a symmetry $\lambda\to \lambda^*$. If we assume the number of unit cells $N$ is even, then consider the two diagonal matrices $\Gamma = {\rm diag}(1,-1,-1,1, \cdots, 1,-1,-1,1)$ and $\Lambda={\rm diag}(1, \sigma_{1,A},\sigma_{2,A},1,\cdots, 1,\sigma_{N-1,A},\sigma_{N,A},1)$. The transformation $\Gamma^{-1}M_L\Gamma=-M_L$ means there is a symmetry $\lambda\to -\lambda$, and the transformed matrix $\Lambda^{-1}M_L\Lambda$ is comprised of entries that are functions of products of two random signs $\sigma_i\sigma_j$, which means there is a statistical symmetry under a rotation of $\pi/2$ in the complex plane: $\lambda\to i\lambda$.

From Figures \ref{fig:ladder_rs_spectra_1} to \ref{fig:ladder_rs_spectra_3}, we see that the delocalization of states caused by increasing the directional bias parameter $g$ occurs in a manner distinct from that of the SSH chain. Initially, when $g=0$, all states are localized, similar to the SSH chain. However, in contrast to the SSH chain, once we increase $g$ and states start to delocalize into loops of extended states, they do not open a hole in the spectrum, as there are localized states still inside the loops. But once $g$ is increased beyond the diabolic value $g_c$ from Eq. (\ref{eq:ladder_gc}), a second hole of extended states opens up from the origin. This results in a spectrum with two loops of extended states, with localized states sandwiched in between. 

\begin{figure} [htbp]
    \centering
    \includegraphics[width=\linewidth]{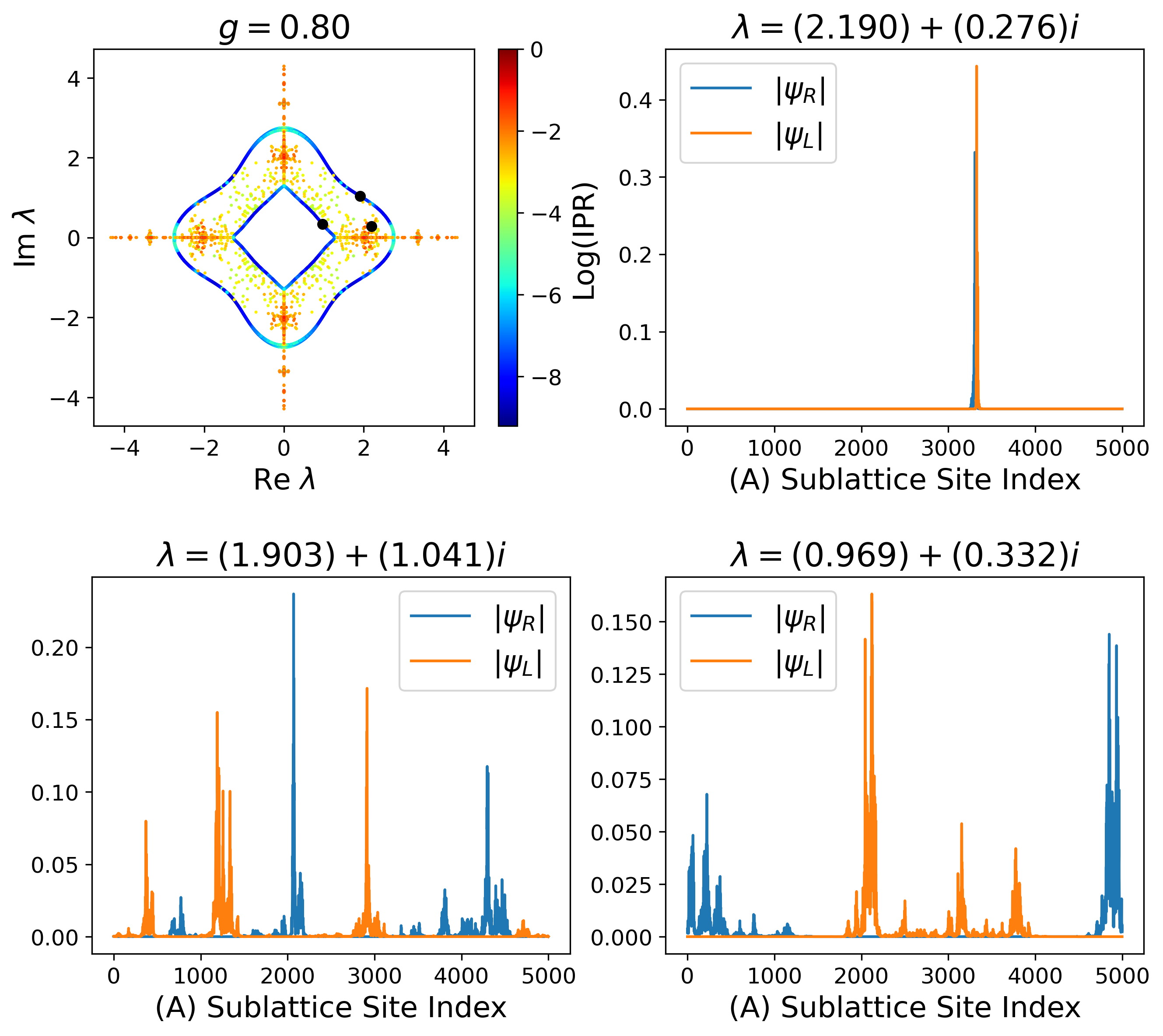}
    \caption{Three representative states for the ladder model with directional bias $g=0.8$ from Figure \ref{fig:ladder_rs_spectra_3}. (Top left) Spectrum of eigenvalues, with the eigenvalues for the three specific eigenfunctions labeled with black dots. (Top right) Right and left eigenfunctions for the localized state at $\lambda=2.190+0.276i$. (Bottom left) Right and left eigenfunctions for the outer loop extended state at $\lambda=1.903+1.041i$. As one traverses the ladder, one alternates three times from regions where the left eigenstate (orange) is dominant to regions where the right eigenstate (blue) is dominant. (Bottom right) Right and left eigenfunctions for the inner loop extended state at $\lambda=0.969+0.332i$. As the ladder is traversed one alternates from regions where the left eigenstate (orange) is dominant to regions where the right eigenstate (blue) is dominant only once.}
    \label{fig:ladder_eigenstates}
\end{figure}

In Figure \ref{fig:ladder_eigenstates}, we show examples of three representative eigenstates for $g=0.8$, with one extended state from the outer loop, another extended state from the inner loop, and one localized state in between. When plotting the eigenstates, we show only the values of the eigenstate norms on the (A) sublattice, because the values are nearly identical on the (B) sublattice. In Figure \ref{fig:ladder_eigenstates}, the top left plot shows the eigenspectrum, with three black circles to mark the eigenvalues corresponding to the states shown in the remaining three plots. The top right plot shows a localized state with eigenvalue $\lambda=2.190+0.276i$. The bottom left plot shows an extended state with eigenvalue $\lambda=1.903+1.041i$ residing in the outer loop, and the bottom right plot shows an extended state with eigenvalue $\lambda=0.969+0.332i$ residing in the inner loop. In each plot, the left and right eigenstates are shown, in orange and blue, respectively. For both extended states, there is little overlap between the left and right eigenstates. However, there is one consistent difference between the extended states in the outer loop versus the inner loop. In the bottom left plot, as one traverses the ladder, the eigenstates alternate from being dominated by its left eigenstate (orange) to its right eigenstate (blue) three times, whereas in the bottom right plot, this only happens once. This difference is consistent across a variety of extended states sampled from the outer and inner loops. 

\subsection{Transfer matrix method for the ladder}

To further investigate the presence of the two distinct bands of extended states that occur in the ladder model, we also use a transfer matrix method. Similar to the SSH chain, the eigenvalue problem $M_L\vec{\psi}=\lambda\vec{\psi}$ in Eq. (\ref{eq:rs_ladder}) can be recast into a system of recursive equations:
\begin{equation}
    te^{-g}\sigma_{j-1,A}\psi_{j-1,A} + u\sigma_{j,B}\psi_{j,B} + te^g\sigma_{j+1,A}\psi_{j+1,A} = \lambda\psi_{j,A},
\end{equation}
\begin{equation}
    te^{-g}\sigma_{j-1,B}\psi_{j-1,B} + u\sigma_{j,A}\psi_{j,A} + te^g\sigma_{j+1,B}\psi_{j+1,B} = \lambda\psi_{j,B}.
\end{equation}
For the ladder model, it is helpful to collect these recursion relations into a $4\times 4$ transfer matrix $T_j$
\begin{equation}
    T_j = \begin{pmatrix}
        \mathbf{0} & \mathbf{1} \\
        Y_1 & Y_2
    \end{pmatrix},
\end{equation}
where $\mathbf{0}$ is the $2\times 2$ matrix of zeros, $\mathbf{1}$ is the $2\times 2$ identity matrix, and $Y_1, Y_2$ are the $2\times 2$ matrices
\begin{equation}
        Y_1 = \begin{pmatrix}
            -e^{-2g}\frac{\sigma_{j-1,A}}{\sigma_{j+1,A}} & 0 \\
            0 & -e^{-2g}\frac{\sigma_{j-1,B}}{\sigma_{j+1,B}}
        \end{pmatrix},
\end{equation}
\begin{equation}
Y_2 = \begin{pmatrix}
            \frac{\lambda}{te^g\sigma_{j+1,A}} & -\frac{u\sigma_{j,B}}{te^g\sigma_{j+1,A}} \\
            -\frac{u\sigma_{j,A}}{te^g\sigma_{j+1,B}} & \frac{\lambda}{te^g\sigma_{j+1,B}}
        \end{pmatrix} .
\end{equation}
The transfer matrix $T_j$ relates the amplitudes of eigenstates across different unit cells:
\begin{equation}
    T_j\begin{pmatrix}
        \psi_{j-1,A} \\ \psi_{j-1, B} \\ \psi_{j, A} \\ \psi_{j, B}
    \end{pmatrix} = 
    \begin{pmatrix}
        \psi_{j,A} \\ \psi_{j, B} \\ \psi_{j+1, A} \\ \psi_{j+1, B}
    \end{pmatrix}.
\end{equation}
If we consider products of $N$ random transfer matrices $M_N=T_NT_{N-1}\cdots T_2T_1$, the Lyapunov exponents $\{\gamma_i\}$ are given by the average logarithmic growth of the singular values $\{s_i\}$ of the product matrix $M_N$ for $N\to\infty$:
\begin{equation}
    \gamma_i=\lim_{N\to\infty}\frac{1}{N}\log s_i(M_N).
\end{equation}
\begin{figure}
    \centering
    \includegraphics[width=0.9\linewidth]{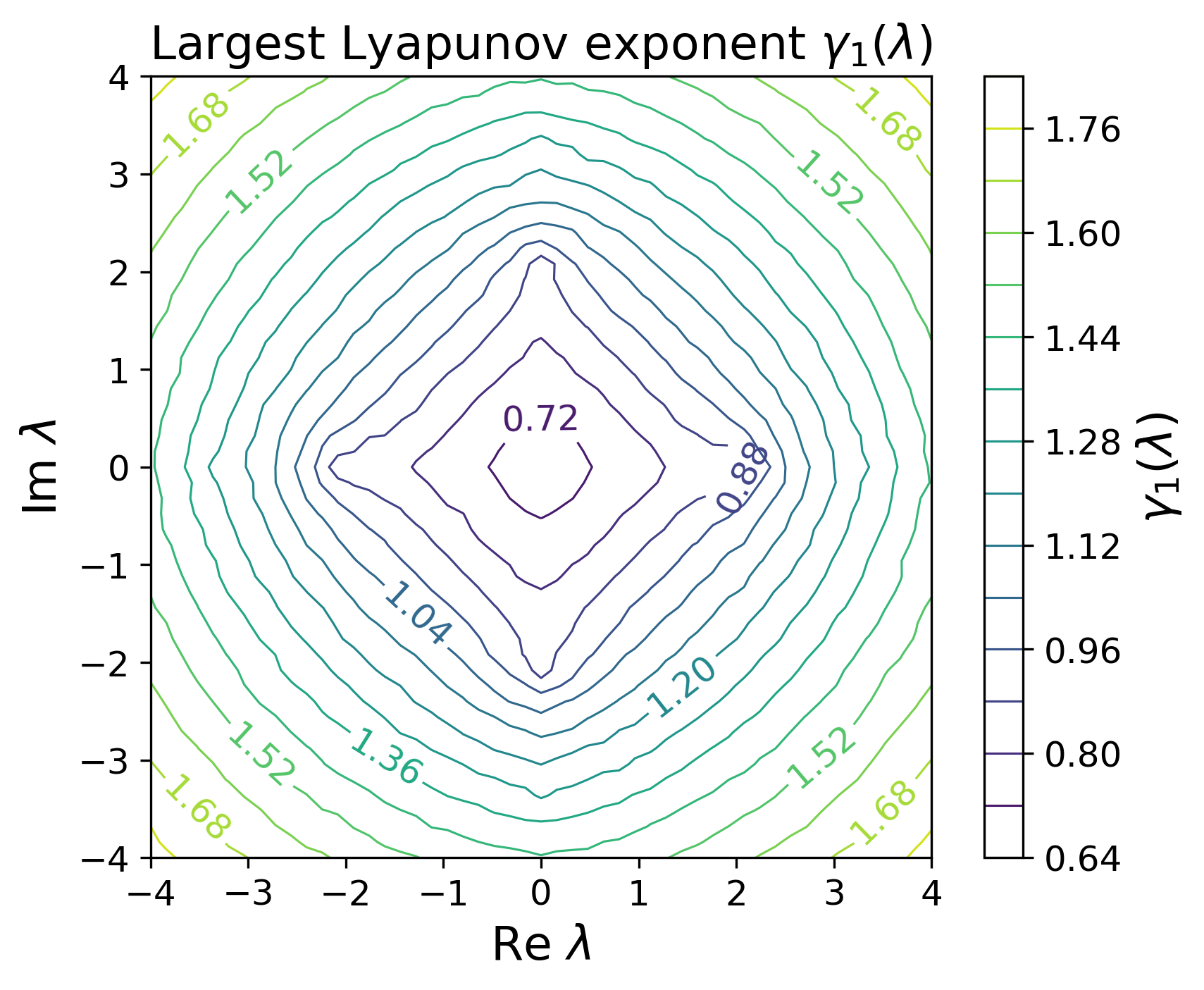}
    \includegraphics[width=0.9\linewidth]{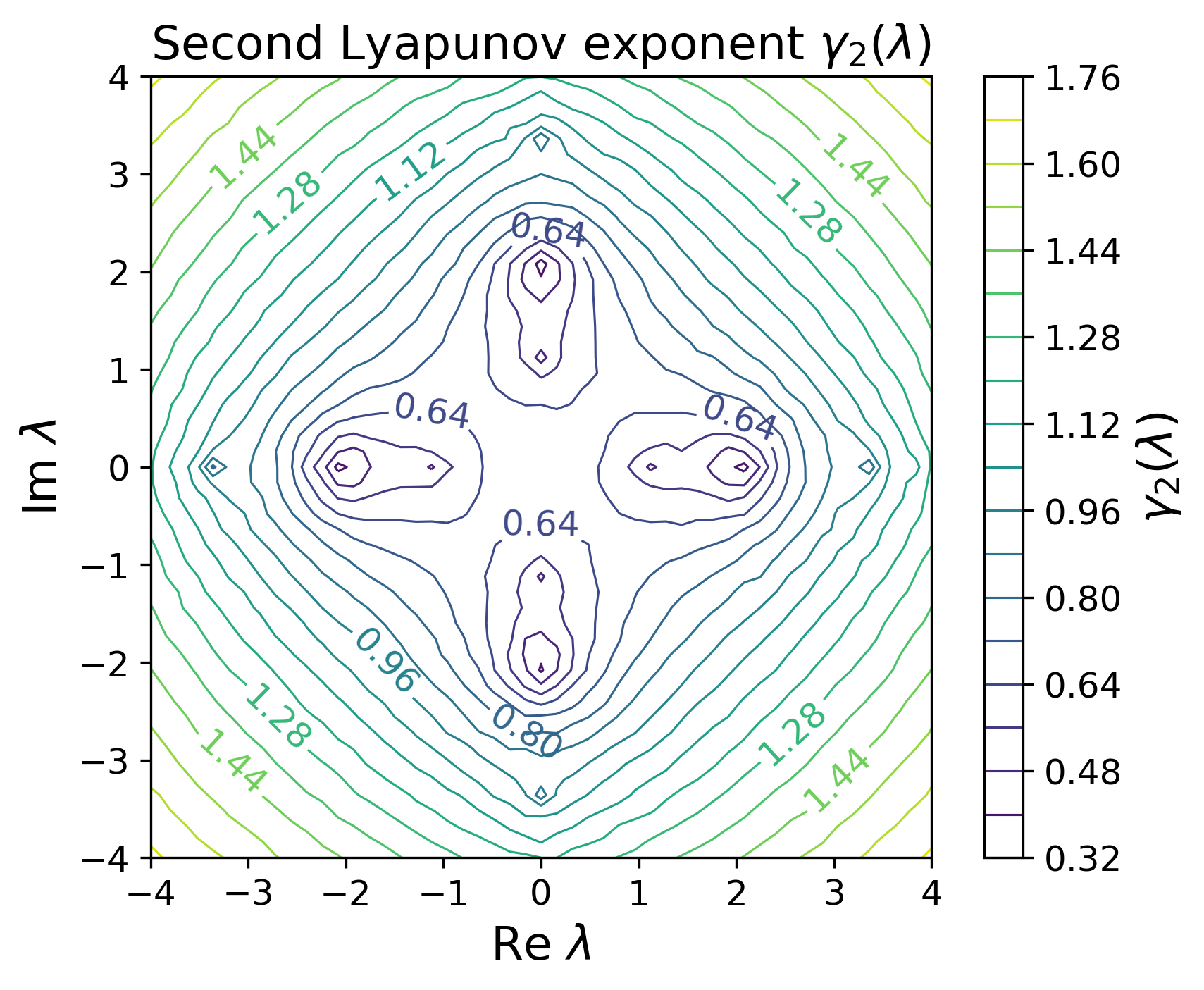}
    \caption{Contour plots of the two Lyapunov exponents $\gamma_1(\lambda)>\gamma_2(\lambda)$ as a function of a complex eigenvalue $\lambda$ for the random sign ladder with $t=1, u=2.5$, and $g=0$. The larger Lyapunov exponent $\gamma_1$ is shown in the top plot, and the smaller Lyapunov exponent $\gamma_2$ is shown in the bottom plot.}
    \label{fig:ladder_lyapunov_contours}
\end{figure}
The Lyapunov exponents can be obtained numerically using methods from \cite{benettin1980theory, benettin1980numerics}. For the ladder model, there are two positive Lyapunov exponents, which we call $\gamma_1>\gamma_2$. These two sets of Lyapunov exponents will correspond to the two loops of extended states seen from direct numerical diagonalization. In Figure \ref{fig:ladder_lyapunov_contours}, we computed the Lyapunov exponents $\gamma_1(\lambda)$ and $\gamma_2(\lambda)$ for $g=0$ over the complex plane, using $N=5000$. The largest Lyapunov exponent $\gamma_1$ is shown in the top plot. It takes on its minimum value at the origin of the complex plane: $\gamma_1(\lambda=0)=g_c$, where $g_c\approx 0.693$ is the diabolic value of $g$ from Eq. (\ref{eq:ladder_gc}). On the other hand, the smaller Lyapunov exponent $\gamma_2$ is shown in the bottom plot. It takes on its minimum value within the bands at values around $\pm 2$ and $\pm 2i$. Previously, in Figures \ref{fig:ladder_rs_spectra_1} to \ref{fig:ladder_rs_spectra_3}, we saw that the delocalization of states occurred in two stages. The first stage occurs for $g<g_c$, when the extended states reside on the contour $g=\gamma_2$. For $g>g_c$, a second set of extended states reside on the contour $g=\gamma_1$. 
\begin{figure}
    \centering
    \includegraphics[width=0.9\linewidth]{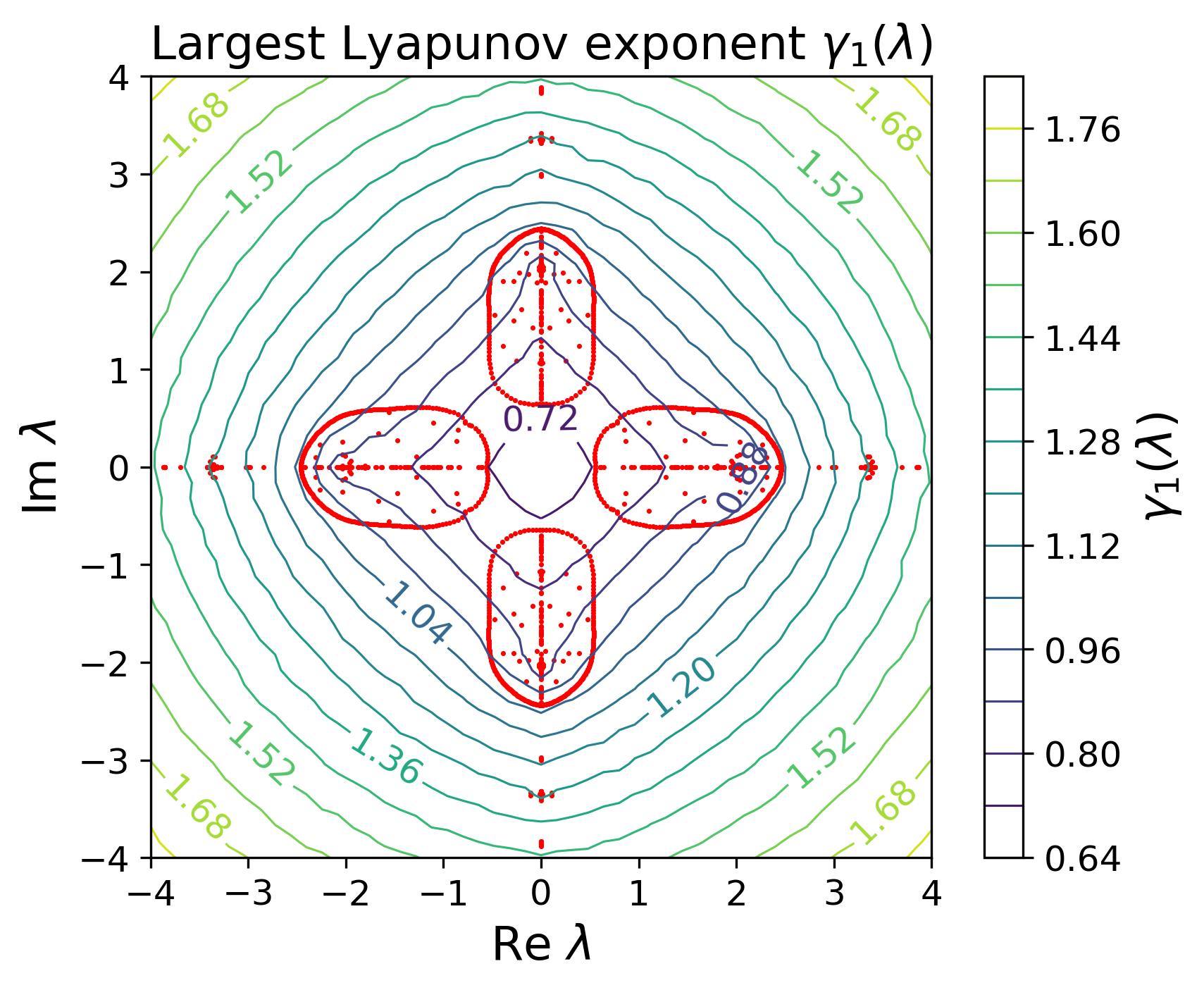}
    \includegraphics[width=0.9\linewidth]{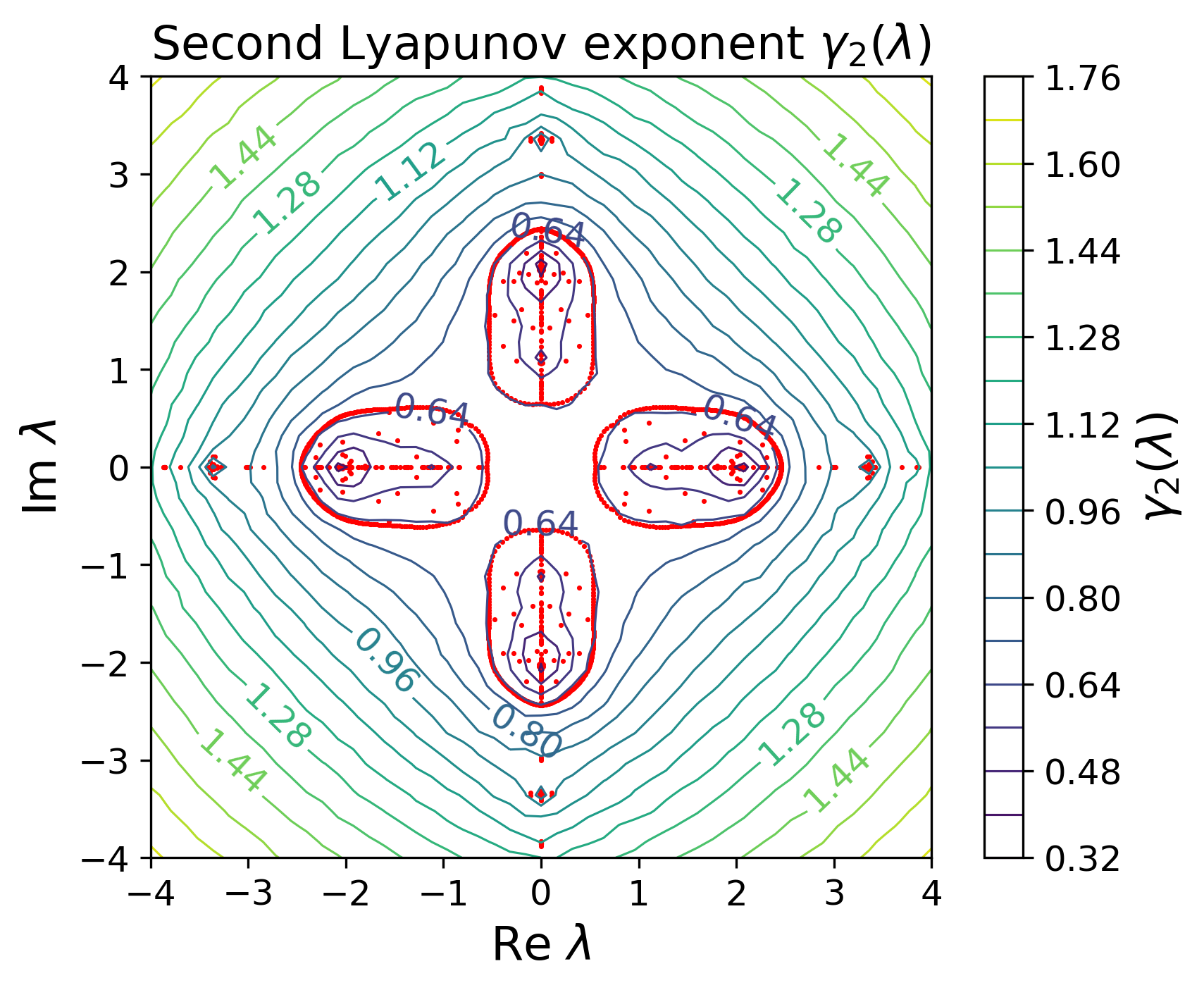}
    \caption{Superimposed on the contours is the eigenspectrum for a particular instance of disorder ($N=1000$ unit cells, $t=1, u=2.5$, and $g=0.64$). Since this is the case where $g<g_c$, the extended states reside on the $\gamma_2=0.64$ contour.}
    \label{fig:ladder_lyapunov_spectra_g_0_64}
\end{figure}
\begin{figure}
    \centering
    \includegraphics[width=0.9\linewidth]{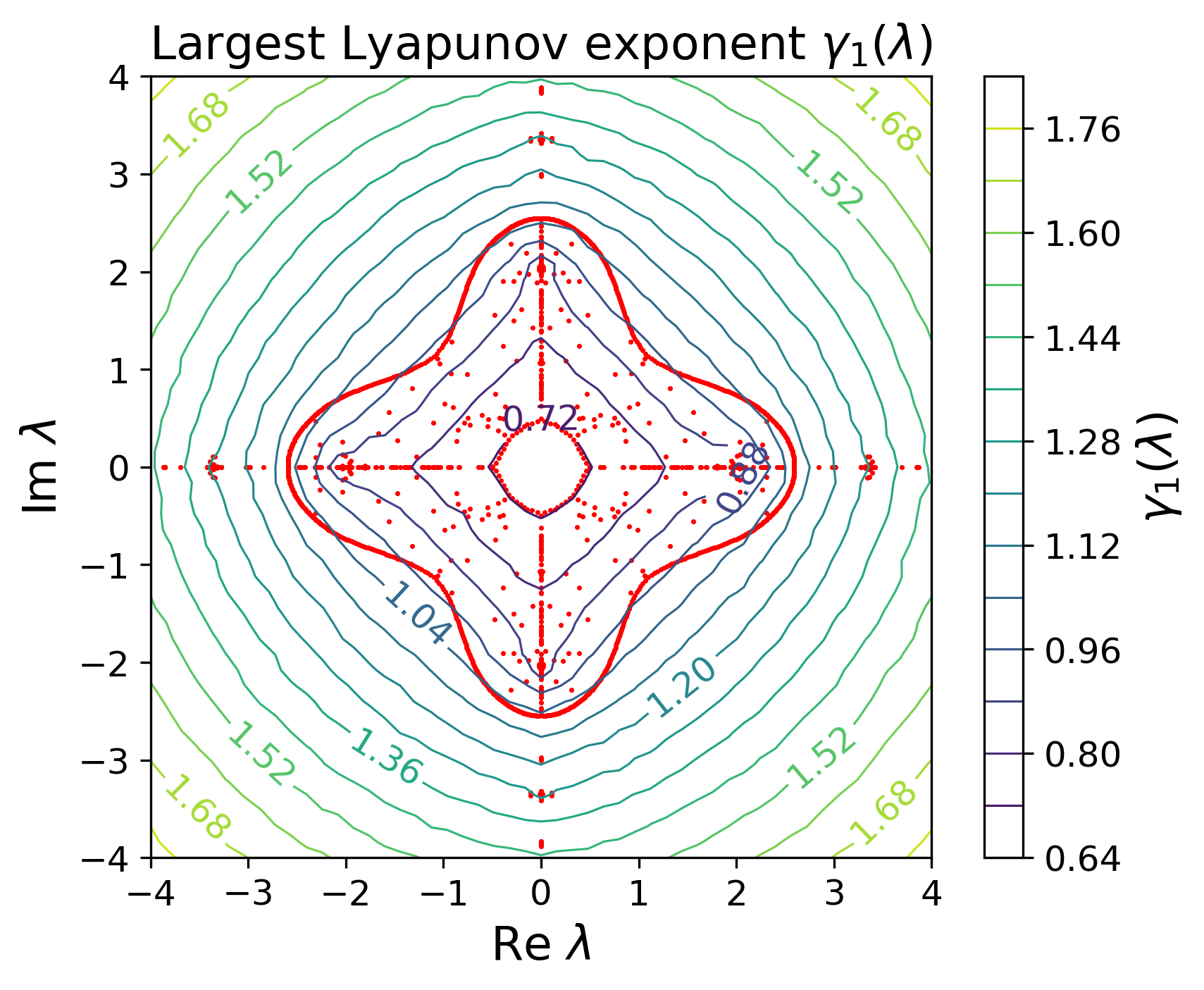}
    \includegraphics[width=0.9\linewidth]{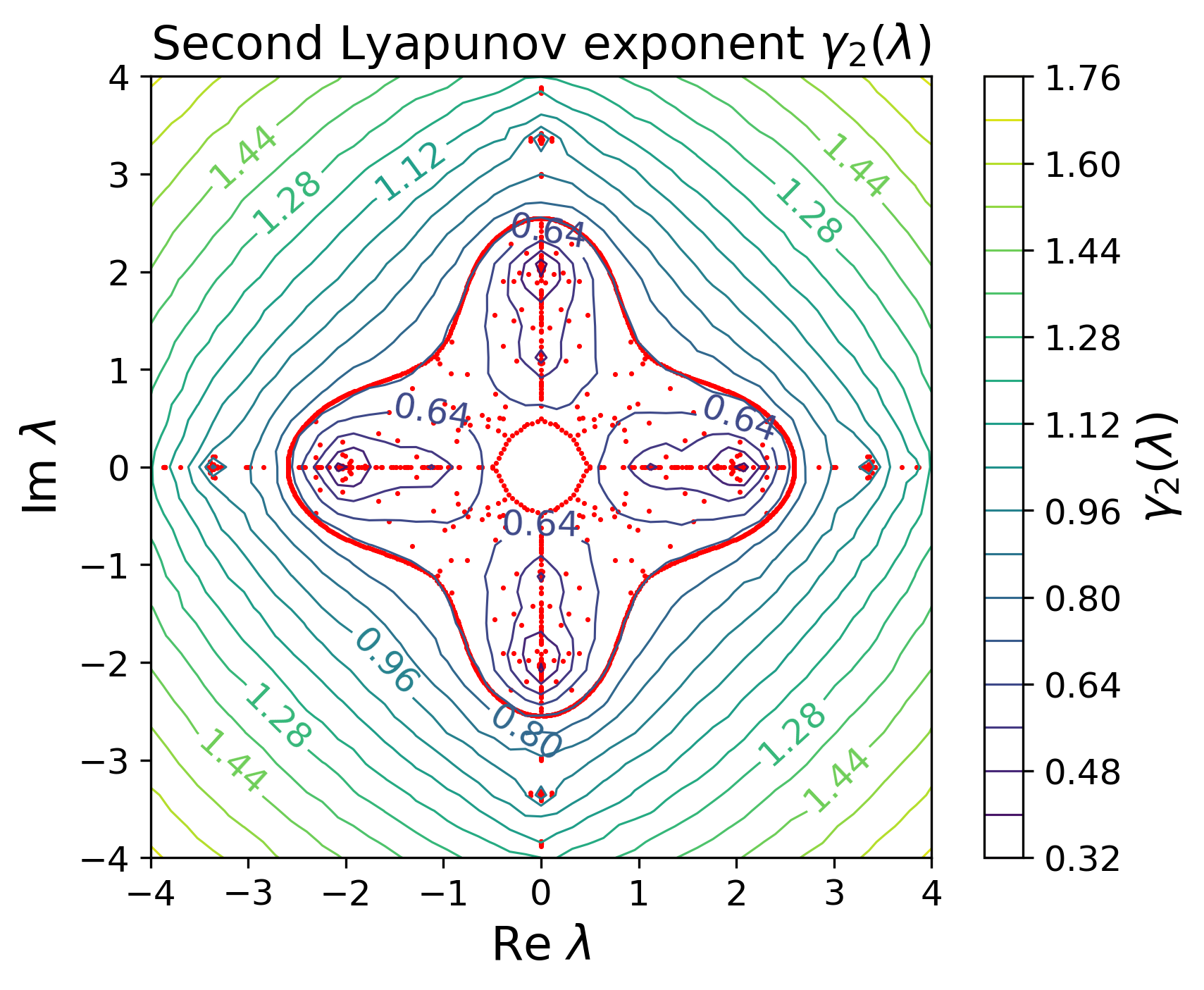}
    \caption{Superimposed on the contours is the eigenspectrum for a particular instance of disorder ($N=1000$ unit cells, $t=1, u=2.5$, and $g=0.72$). Since this is the case where $g>g_c$, there are two sets of extended states: the outer set which lies on the $\gamma_2=0.72$ contour, and the inner set which lies on the $\gamma_1=0.72$ contour.}
    \label{fig:ladder_lyapunov_spectra_g_0_72}
\end{figure}

In Figures \ref{fig:ladder_lyapunov_spectra_g_0_64} and \ref{fig:ladder_lyapunov_spectra_g_0_72}, we superimpose particular instances of random ladder eigenspectra for the two cases of $g<g_c$ and $g>g_c$. In Figure \ref{fig:ladder_lyapunov_spectra_g_0_64}, we consider the case $g<g_c$ and find that the extended states indeed lie on the $\gamma_2=g$ contour. On the other hand, in Figure \ref{fig:ladder_lyapunov_spectra_g_0_72}, we consider the case $g>g_c$ and find that the extended states indeed lie on the two sets of contours $\gamma_2=g$ and $\gamma_1=g$.

\section{Open boundary conditions} \label{sec:obc}

Compared to their Hermitian counterparts, non-Hermitian systems have an enhanced sensitivity to boundary conditions \cite{dahmen2007population}. Under periodic boundary conditions, the directional bias $g$ tends to delocalize the eigenstates. However, under open boundary conditions, the directional bias leads to an accumulation of states at the edge, known as the non-Hermitian skin effect \cite{yao2018edge, wang2024non}. We can understand the effects of open boundary conditions on both the eigenspectrum and eigenstates by considering appropriate gauge transformations for the ladder and SSH chain. In the case of the ladder model, open boundaries mean the upper and lower corner entries of the banded connectivity matrix $M_L$ in Eq. (\ref{eq:rs_ladder_matrix}) are set to zero:
\begin{equation}
    M_L' = \begin{pmatrix}
        R_1 & L_2^+ & & & \\
        L_1^- & R_2 & L_3^+ \\
        & L_2^- & \ddots & \ddots \\
        & & \ddots & \ddots & L_N^+ \\
         & & & L_{N-1}^- & R_N
    \end{pmatrix},
\end{equation}
where the $2\times 2$ block matrices $R_j$ and $L_j^\pm$ remain defined as in Eq. (\ref{eq:rs_ladder_blocks}). The absence of corner matrix elements means we can gauge out the dependence on $g$ by using the diagonal matrix $\Omega_L={\rm diag}(1,1,e^{-g},e^{-g},\cdots,e^{-(N-1)g},e^{-(N-1)g})$:
\begin{equation}
    \Omega_L^{-1}M_L'\Omega_L = \begin{pmatrix}
        R_1 & L_2 & & & \\
        L_1 & R_2 & L_3 \\
        & L_2 & \ddots & \ddots \\
        & & \ddots & \ddots & L_N \\
         & & & L_{N-1} & R_N
    \end{pmatrix},
\end{equation}
where the $R_j$ blocks remain the same but the off-diagonal blocks $L_j$ are now independent of the directional bias $g$:
\begin{equation}
    L_j=t\begin{pmatrix}
        \sigma_{j,A} & 0 \\
        0 & \sigma_{j,B}
    \end{pmatrix}.
\end{equation}
Thus, under open boundary conditions, even after turning on the directional bias $g$, the eigenvalues of the ladder model remain the same as in the $g=0$ case. In addition, we note that the ladder eigenspectrum for $g=0$ is roughly the same whether under open or periodic boundaries.

While the directional bias has no effect on the eigenvalues for the open boundary ladder, it can dramatically change the eigenstates \cite{dahmen2007population}. If an eigenvalue $\lambda$ has the right eigenvector $\vec{\psi}$ when $g=0$, then $\Omega_L\vec{\psi}$ is the corresponding right eigenvector for $g\neq 0$. As a result, right eigenstates localize or pile up near the left boundary for $g>0$. The left eigenstates pile up near the right boundary when $g>0$ for similar reasons.

The same phenomenon occurs for the SSH chain. For open boundary conditions, the connectivity matrix $M_S$ for the random sign SSH chain in Eq. (\ref{eq:rs_ssh_matrix}) is modified to 
\begin{equation}
    M_S' = \begin{pmatrix}
        & p_2 & & & \\
        \ell_1 & & p_3 \\
        & \ell_2 & & \ddots \\
        & & \ddots & & p_{2N} \\
         & & & \ell_{2N-1}
    \end{pmatrix},
\end{equation}
with the only change occurring in the upper right and lower left corner entries, which are now set to zero. For the open SSH chain, we use the diagonal gauge transformation $\Omega_S={\rm diag}(1,e^{-g},e^{-2g},\cdots, e^{-(2N-1)g})$ to get
\begin{equation}
    \Omega_S^{-1}M_S'\Omega_S = \begin{pmatrix}
        & p_2' & & & \\
        \ell_1' & & p_3' \\
        & \ell_2' & & \ddots \\
        & & \ddots & & p_{2N}' \\
         & & & \ell_{2N-1}'
    \end{pmatrix},
\end{equation}
where the non-zero entries are 
\begin{equation}
    \begin{split}
        \ell_i' &= \begin{cases}
        t^+\sigma_{\frac{i+1}{2}, A} \quad &(\text{for $i$ odd})\\
        t^-\sigma_{\frac{i}{2}, B} \quad &(\text{for $i$ even}),
    \end{cases} \\
    p_i' &= \begin{cases}
        t^-\sigma_{\frac{i+1}{2}, A} \quad &(\text{for $i$ odd})\\
        t^+\sigma_{\frac{i}{2}, B} \quad &(\text{for $i$ even}),
    \end{cases}
    \end{split}
\end{equation}
showing that under open boundaries, the SSH chain also has eigenvalues independent of the directional bias $g$, with right eigenstates localizing towards the left boundary and left eigenstates piling up on the right boundary as $g>0$.

\begin{figure}
    \centering
    \includegraphics[width=\linewidth]{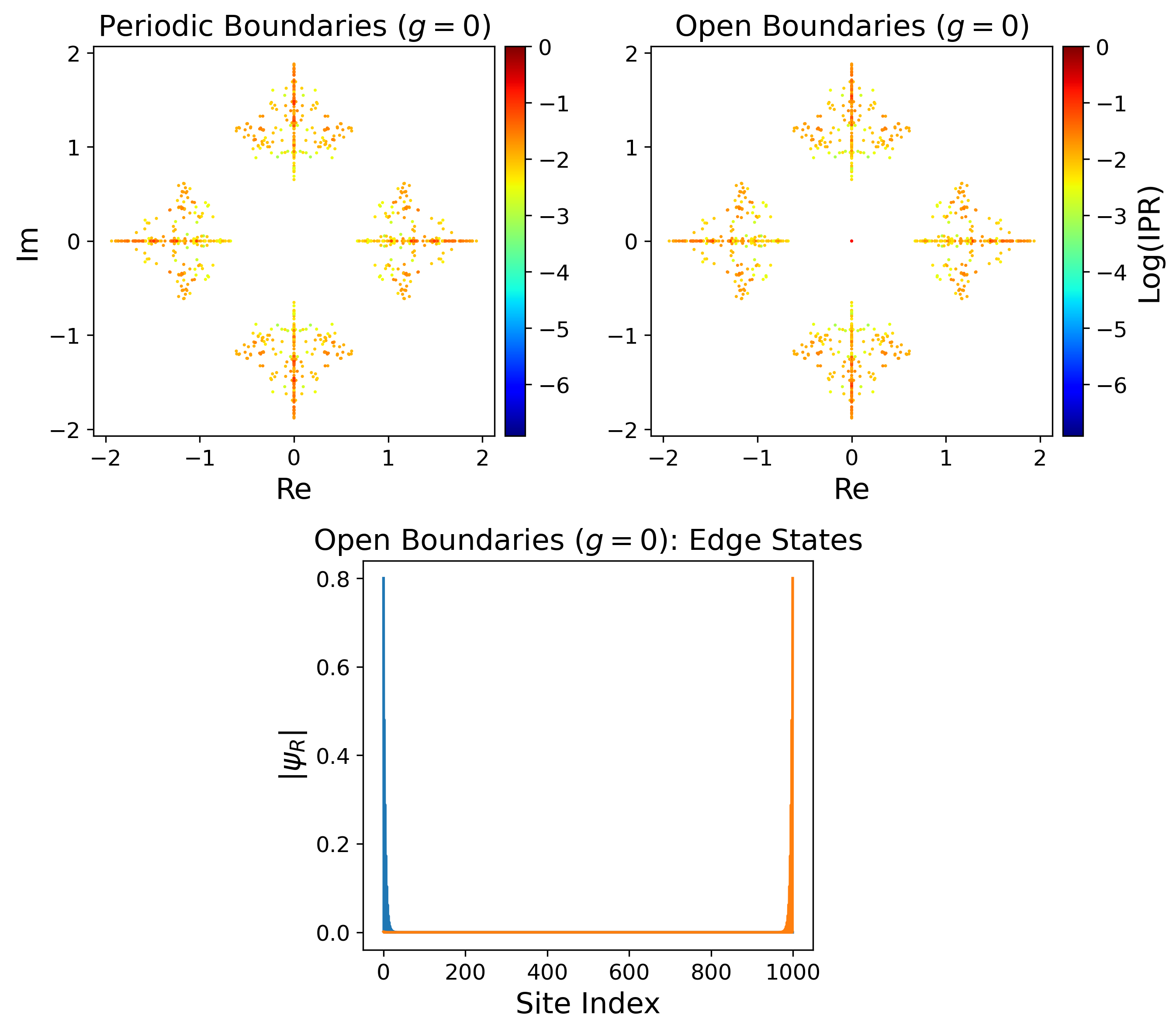}
    \caption{The random-sign SSH chain for $N=500$ unit cells, with no directional bias $(g=0)$ and hopping parameters $t^+=0.75$ and $t^-=1.25$. The two plots in the top row show the eigenspectra for periodic (left) and open (right) boundaries. The system under open boundaries has two edge states with eigenvalues near zero (depicted in the last plot).}
    \label{fig:ssh_obc}
\end{figure}

However, there is one distinction for the SSH chain compared to the ladder. In the absence of any directional bias ($g=0$), the ladder eigenspectra for periodic and open boundaries are the same for large $N$. For the SSH chain, this is also true if the two hopping parameters satisfy $t^+>t^-$. On the other hand, if $t^+<t^-$, then the open boundary disordered SSH chain has two edge states (with near zero eigenvalues) that do not appear for periodic boundaries, reminiscent of the SSH chain without disorder. In Figure \ref{fig:ssh_obc}, we illustrate this phenomenon for $N=500$ with $t^+=0.75$ and $t^-=1.25$, and we see that these edge modes in fact persist in the non-Hermitian, random-sign case.

\section{Discussion} \label{sec:discussion}

In this paper, inspired by ring neural networks, we studied the localization properties of multi-banded random matrices with directional bias and random sign disorder in the hoppings, consistent with Dale's Law. Unlike many non-Hermitian models considered thus far \cite{ashida2020non, bergholtz2021exceptional}, asymmetric nearest neighbor couplings inspired by neural networks can be of opposite sign in the two directions \cite{amir2016non}, as opposed to merely having different strengths \cite{hatano1996localization}. We focused on two paradigmatic non-Hermitian systems: the SSH chain and a ladder model. Even without disorder, these two-band systems can have an eigenvalue degeneracy when the directional bias $g$ is tuned to special values. In the case of the SSH chain, the degeneracy occurs via an exceptional point, while in the case of the ladder, a diabolic point appears instead. In both cases, it was found these special points could survive the introduction of random sign disorder. When \textit{both} random sign disorder and directional bias $g$ are present, the delocalization properties of the two models are qualitatively different. While the delocalization of states for the SSH chain was similar to that of earlier one banded models, for the ladder model, the delocalization occurs in two distinct stages, controlled by whether the directional bias $g$ is greater or less than its diabolic value. This results in two distinct contours of extended states, with localized states in between. The contours of extended states can be understood for both models in terms of the Lyapunov exponents of transfer matrices, in agreement with direct diagonalization results. Finally, when under open boundary conditions, both systems show a non-Hermitian skin effect, with states localizing at a boundary. For the open SSH chain in particular, it is also possible for the system to have edge modes, similar to the topologically non-trivial phase for the pure SSH chain \cite{wang2024non}. 

The results on random eigenvalues and the localization properties of their eigenstates have direct relevance for neural network dynamics, as the stability of the linearized firing rate dynamics in Eq. (\ref{eq:firing_rates}) is determined by the eigenvalues of the network connectivity matrix $M_{S/L}$. In our models, the largest eigenvalue real parts can be driven positive, despite the $-\mathbf{v}$ term in Eq. (\ref{eq:firing_rates}); this leads to unstable dynamics, which correspond to spontaneous activity in the network \cite{vogels2005neural}. While eigenvalues with positive real parts can arise for all values of the directional bias $g$, the bias determines whether the most unstable states are localized. For small bias, this means the spontaneous activity occurs in localized regions of the network. On the other hand, a large enough bias will not only increase the magnitudes of the most unstable eigenvalues, but also delocalize all the states, as indicated by both the increasing radii of the singular value annular bounds as well as the Lyapunov exponent level curves. In this case, spontaneous activity occurs in a spatially extended way across the network. While we have focused on the localization properties of disordered two-banded models, it would be interesting to study disordered networks with a more complex spatial substructure or with unit cells replaced by neural clusters with dense, random internal connections.

\section{Acknowledgments}

We thank L. Mahadevan for helpful discussions, and D.R.N. acknowledges useful discussions with Grace Zhang during the early part of this work. This work was supported by the National Science Foundation, through the Harvard University Materials Research Science and Engineering Center, Grant No. DMR-2011754.

\newpage

\bibliographystyle{unsrt}
\bibliography{refs}

\end{document}